\documentclass[twocolumn]{aastex6}
\usepackage{amsmath,amstext}
\usepackage[T1]{fontenc}
\usepackage{apjfonts}
\usepackage[figure,figure*]{hypcap}
\definecolor{xlinkcolor}{cmyk}{1,1,0,0}
\hypersetup{colorlinks, linkcolor=xlinkcolor, anchorcolor=xlinkcolor, citecolor=xlinkcolor}
\usepackage{bm}

\begin{document}

\title{Exploring The origin of multiwavelength activities of high-redshift FSRQ PKS~1502+106 during 2014--2018}

\author{N. Ding\altaffilmark{1,2}, Q. S. Gu\altaffilmark{1,2}, X. F. Geng\altaffilmark{3,4}, DING-RONG Xiong\altaffilmark{5,6}, R. Xue\altaffilmark{1,2}, X. Y. Wang\altaffilmark{1,2}, X. T. Guo\altaffilmark{1,2}}
\altaffiltext{1}{School of Astronomy and Space Science, Nanjing University, Nanjing, Jiangsu 210093, China}
\altaffiltext{2}{Key Laboratory of Modern Astronomy and Astrophysics (Nanjing University), Ministry of Education, Nanjing 210093, China}
\altaffiltext{3}{Department of Physics, Yunnan University, Kunming, 650091, China}
\altaffiltext{4}{Key Laboratory of Astroparticle Physics, Yunnan Province, Kunming 650091, China}
\altaffiltext{5}{Yunnan Observatories, Chinese Academy of Sciences, Kunming 650216, China}
\altaffiltext{6}{Key Laboratory for the Structure and Evolution of Celestial Objects, Chinese Academy of Sciences, Kunming 650216, China}


\begin{abstract}
The origin of the multi-band activities (outbursts/flares) of blazars is still a heavily debated topic. Shock and magnetic reconnection have long been considered as possible triggers for the multi-band activities. In this paper, we present an exploration of the origin of multi-band activities for a high-redshift (z =1.8385) FSRQ PKS~1502+106. Utilizing multi-band data from radio to $\gamma$-ray and optical polarization observations, we investigate two dramatic activities in detail: a $\gamma$-ray dominated outburst in 2015 and an optical dominated outburst in 2017. Our main results are as follows. 
(I) A fast $\gamma$-ray flare with a flux-doubling time-scale as short as 1-hr in 2015 is discovered. Based on the variability time-scale, the physical parameters of the flaring region (e.g, minimum Doppler factor, emission region size, etc.) are constrained. At the peak of the flare, the $\gamma$-ray spectrum hardens to \mbox{$\Gamma_{\gamma} = 1.82\pm0.04$} and exhibits an obvious curvature/break characteristic that is caused by the typical "cooling break". 
Modelings of multi-band SEDs reveal a very hard electronic energy spectrum with the electronic spectral index of $1.07\pm0.53$. 
This result suggests that this fast $\gamma$-ray flare may be triggered by magnetic reconnection.
(II) During the outburst in 2017, the optical polarization degree and optical fluxes show a very tight correlation. By analyzing Stokes parameters of polarization observations, our results show that this outburst could be triggered by a transverse shock with a compression ratio of $\eta> 2.2$, and the magnetic field intensity of the shock emission region is about $0.032$~G.
\end{abstract}


\maketitle

\section{INTRODUCTION}
Blazars, including BL Lac objects and flat-spectrum radio quasars (FSRQs), are the most extreme type of radio-loud active galactic nuclei (AGN); their relativistic jets point close to our line of sight, for which they possess distinctive observational characteristics, such as large amplitude and rapid variability, high and variable polarization, and compact radio emission \citep[see, e.g.,][]{1995PASP..107..803U, 2015A&ARv..24....2M, 2017A&ARv..25....2P}. Their radiation spans the entire electromagnetic wave, and as \mbox{non-catastrophic} sources, they can be monitored for a long time, thus becoming one of the unique objectives in the era of multi-messenger astronomy \citep[see, e.g.,][]{2019arXiv190304504R, 2019arXiv190304461B}.

Multi-band variability in blazars is an important window to understand the non-thermal radiation mechanisms and the energy dissipation of relativistic jets. Generally, the variability of blazars exhibits random behavior \citep[e.g.,][]{2010ApJ...722..520A, 2012ApJ...749..191C, 2017A&A...598A..39H}, similar to the variability of radio-quiet AGNs \citep[e.g.,][]{2003ApJ...598..935M}. However, in some activities (outbursts/flares), the multi-band variability of blazars presents recognizable patterns \citep[e.g.,][]{2008Natur.452..966M}. These activities possess a variety of characteristics, with variability time-scales ranging from minutes to years \citep[see, e.g.,][]{2013MNRAS.436.1530R, 2016ApJ...824L..20A, 2018ApJ...854L..26S}. Using the shortest variability time-scale of these activities, we can constrain the physical properties of the emission region, such as the size, the magnetic field intensity, and the location of the high-energy emission region \citep[e.g.,][]{2004ApJ...609..576B, 2018ApJ...859..168Y, 2018ApJ...853...34Z}. However, the physical origin of these activities is still not well understood. A variety of scenarios have been proposed (see \citealt{2019Galax...7...20B} for a review), and these scenarios can be roughly classified into models associated with shock (i.e., shock-in-jet models, e.g., \citealt[][]{1985ApJ...298..114M, 2010ApJ...711..445B}), models related to turbulence/magnetic reconnection \citep[e.g.,][]{2009MNRAS.395L..29G, 2013MNRAS.431..355G, 2015MNRAS.450..183S}, models in which jets interact with external environments (e.g., jet-star collision model; \citealt{2012ApJ...749..119B, 2013MNRAS.436.3626A}), and geometric models \citep[e.g.,][]{2013ApJ...768...40L, 2017Natur.552..374R}. It can be noted that shock and magnetic reconnection are the two main internal causes triggering these multi-band activities.

PKS~1502+106 is a powerful FSRQ at a high redshift of $z=1.8385$ \citep[][]{2008ApJS..175..297A}. \mbox{High-resolution} \mbox{very-long-baseline} interferometry (VLBI) observations at mm wavelengths reveal that it possesses a compact, \mbox{core-dominated} morphology with a one-sided, bent, magnetically dominated parsec-scale jet; the jet has a fast superluminal motion with the estimated Doppler factors ranging from $\sim7$ up to $\sim50$ along the jet, showing obvious acceleration characteristic \citep[][]{2016A&A...586A..60K}. At the beginning of the \emph{Fermi} satellite operation (August 2008), PKS~1502+106 exhibited a sudden high-energy flare, becoming the second brightest extragalactic source in the $\gamma$-ray sky at that time. This was followed by high and variable fluxes over the subsequent months, triggering an intensive multi-band campaign covering the radio, optical, ultraviolet (UV), and X-ray bands. \citet[][]{2010ApJ...710..810A} utilized these synergistic observation data to analyze this dramatic outburst. Their results suggest that PKS 1502+106 is a sub-GeV peaked blazar, and its energy dissipation probably occurs within the broad-line region (BLR). 

A minor renewed $\gamma$-ray activity observed by the \emph{Fermi}-LAT was announced again in Jan 2009 through ATel \#1905. After this, PKS~1502+106 entered a quiescent stage of up to six years. Until mid-2015, a highest-level renewed $\gamma$-ray activity was detected by the \emph{Fermi}-LAT (see ATel \#7801), and associated outbursts from near-infrared to UV bands were also observed (see ATel \#7804 and ATel \#7783). Subsequently, in December 2017, Steward Observatory observed extremely high optical polarization from PKS 1502+106. The optical polarization degree is as high as  $(47.4\pm0.1)\%$, which is among the highest levels of polarization ever observed for blazars (see ATel \#11047). Long-term optical photometric monitoring indicates that PKS 1502+106 was undergoing a prominent optical outburst. Motivated by these dramatic multi-band activities, in this paper, we systematically investigate the multiwavelength activities of the source during 2014--2018 using multi-band data from radio to \mbox{$\gamma$-ray} bands as well as optical polarization observations. Especially, two dramatic outbursts, a $\gamma$-ray dominated outburst in 2015 and an optical dominated outburst in 2017, are analyzed in detail to explore the physical origin of their multiwavelength activities and the physical properties of emission regions. 

This paper is organized as follows. Section~2 describes multi-band observations and data reduction. Section~3 presents the global characteristics of the multi-band variability of the source during 2014--2018. The $\gamma$-ray  dominated outburst in 2015 is investigated and discussed in Section~4, and the optical dominated outburst in 2017 is analyzed and discussed in Section~5. In Section~6, we summarize the main results of this work. Cosmological parameters of $H_{0}=70~$km~s$^{-1}$Mpc$^{-1}$, $\Omega_{m}=0.3$, and $\Omega_{\Lambda}=0.7$ \citep[][]{2016A&A...594A..13P} are adopted in this work.

\section{Observations and Data Reduction}
In this section, we present the multiwavelength observations (from radio to $\gamma$-rays bands) of PKS 1502+106 from October 2014 to October 2018 and the processes of data reduction.

\begin{table*}
\tablenum{1}
\begin{center}
\caption{Summary of Swift-XRT Observations of PKS~1502+106}
\footnotesize
\begin{tabular}{@{}lccccc@{}}
\hline\hline
Obs-ID						&	Date (MJD)		&	Net Exp. 			&	Flux					&	$\Gamma_{\mathrm{X}}$		&						C-stat./dof 				\\
							&					&	(s)				&	($10^{-12}$~erg~cm$^{-2}$~s$^{-1}$)					&						&												\\
(1)	&	(2)	&	(3)	&	(4)	&	(5)	 &	 (6)			\\
\hline
00036388020$^{a}$	&	2015-07-08	(57,211.63)	&	4722.40	&	1.29	$\pm$	0.16	&	$1.85_{-0.13}^{+0.13} $	 &	 119.4 (137)			\\
00036388021$^{b}$	&	2015-07-12	(57,215.14)	&	1982.85	&	1.18	$\pm$	0.27	&	$1.70_{-0.25}^{+0.25} $	 &	 44.3  (67)				\\
00081590001$^{c}$		&	2015-07-14	(57,217.74)	&	1376.01	&	1.86	$\pm$	0.51	&	$1.08_{-0.37}^{+0.36} $	 &	 47.8  (46)				\\
00036388022$^{d}$	&	2015-07-15	(57,218.34)	&	2976.76	&	0.41	$\pm$	0.11	&	$1.80_{-0.27}^{+0.30} $	 &	 47.6  (45)				\\
00036388023	&	2015-07-18	(57,221.27)	&	2904.34	&	1.36	$\pm$	0.25	&	$1.40_{-0.19}^{+0.20} $	 &	 69.2  (82)				\\
00036388024	&	2015-07-21	(57,224.13)	&	2881.86	&	1.73	$\pm$	0.33	&	$1.23_{-0.17}^{+0.17} $	 &	 66.3  (92)				\\
00036388025	&	2015-07-24	(57,227.59)	&	1882.97	&	1.54	$\pm$	0.32	&	$1.54_{-0.20}^{+0.20} $	 &	 61.8  (70)				\\
00036388027	&	2016-05-23	(57,531.82)	&	2976.76	&	1.42	$\pm$	0.24	&	$1.44_{-0.18}^{+0.17} $	 &	 79.2  (86)				\\
00092241001	&	2016-06-18	(57,557.09)	&	794.15	&	1.04	$\pm$	0.41	&	$1.80_{-0.41}^{+0.42} $	 &	 12.3  (26)				\\
00092241002	&	2016-06-26	(57,565.45)	&	1038.87	&	0.58	$\pm$	0.18	&	$1.95_{-0.49}^{+0.63} $	 &	 19.4  (25)				\\
00092241003	&	2016-07-04	(57,573.42)	&	891.53	&	0.86	$\pm$	0.65	&	$1.81_{-0.60}^{+0.60} $	 &	 15.5  (24)				\\
00092241004	&	2016-07-12	(57,581.27)	&	1023.90	&	0.71	$\pm$	0.36	&	$1.43_{-0.57}^{+0.57} $	 &	 23.1  (25)				\\
00092241005	&	2016-07-20	(57,589.24)	&	976.44	&	1.25	$\pm$	0.51	&	$1.40_{-0.43}^{+0.42} $	 &	 9.9   (24)				\\
00092241006	&	2016-07-28	(57,597.35)	&	976.44	&	0.92	$\pm$	0.42	&	$1.52_{-0.51}^{+0.52} $	 &	 13.9  (24)				\\
00092241007	&	2016-12-19	(57,741.20)	&	978.94	&	0.81	$\pm$	0.29	&	$1.45_{-0.53}^{+0.61} $	 &	 35.7  (28)				\\
00092241008	&	2016-12-27	(57,749.90)	&	933.99	&	1.56	$\pm$	0.55	&	$1.15_{-0.38}^{+0.37} $	 &	 36.5  (32)				\\
00092241009	&	2017-01-04	(57,757.55)	&	966.45	&	0.96	$\pm$	0.34	&	$1.46_{-0.42}^{+0.41} $	 &	 22.8  (29)				\\
00092241010	&	2017-01-12	(57,765.18)	&	961.46	&	1.55	$\pm$	0.55	&	$1.50_{-0.30}^{+0.31} $	 &	 32.8  (41)				\\
00092241011	&	2017-01-19	(57,772.56)	&	1091.31	&	1.31	$\pm$	0.40	&	$1.50_{-0.33}^{+0.34} $	 &	 31.2  (36)				\\
00092241012	&	2017-01-28	(57,781.13)	&	1068.84	&	0.79	$\pm$	0.21	&	$1.82_{-0.39}^{+0.40} $	 &	 18.6  (30)				\\
00092241014	&	2017-02-08	(57,792.64)	&	566.89	&	1.94	$\pm$	0.83	&	$1.19_{-0.38}^{+0.38} $	 &	 21.2  (26)				\\
00093162001	&	2017-06-23	(57,927.76)	&	1083.82	&	1.24	$\pm$	0.31	&	$1.18_{-0.50}^{+0.59} $	 &	 20.3  (26)				\\
00093162002	&	2017-06-30	(57,934.86)	&	1078.83	&	1.28	$\pm$	0.35	&	$1.68_{-0.40}^{+0.45} $	 &	 22.8  (35)				\\
00093162003	&	2017-07-07	(57,941.38)	&	1098.81	&	1.96	$\pm$	0.63	&	$1.06_{-0.34}^{+0.33} $	 &	 43.8  (44)				\\
00093162004	&	2017-07-14	(57,948.48)	&	1003.91	&	1.72	$\pm$	0.70	&	$1.45_{-0.41}^{+0.39} $	 &	 35.3  (36)				\\
00093162005	&	2017-07-21	(57,955.33)	&	1023.89	&	1.30	$\pm$	0.38	&	$1.84_{-0.33}^{+0.37} $	 &	 29.2  (39)				\\
00093162006	&	2017-07-28	(57,962.04)	&	1048.86	&	1.00	$\pm$	0.32	&	$1.75_{-0.54}^{+0.60} $	 &	 21.1  (29)				\\
00093162007	&	2017-12-22	(58,109.64)	&	966.45	&	1.92	$\pm$	0.46	&	$1.91_{-0.28}^{+0.27} $	 &	 34.3  (53)				\\
00036388028	&	2017-12-27	(58,114.29)	&	2647.14	&	1.67	$\pm$	0.23	&	$1.67_{-0.17}^{+0.18} $	 &	 93.3  (99)				\\
00093162008	&	2017-12-29	(58,116.62)	&	1008.91	&	1.59	$\pm$	0.32	&	$1.69_{-0.28}^{+0.30} $	 &	 37.4  (46)				\\
00036388029	&	2017-12-30	(58,117.02)	&	4729.87	&	2.17	$\pm$	0.19	&	$1.71_{-0.10}^{+0.10} $	 &	 130.5 (180)			\\
00036388030	&	2018-01-02	(58,120.14)	&	3141.60	&	1.60	$\pm$	0.21	&	$1.64_{-0.15}^{+0.15} $	 &	 95.5  (109)			\\
00093094001	&	2018-01-04	(58,122.98)	&	1493.37	&	2.71	$\pm$	0.68	&	$1.04_{-0.22}^{+0.22} $	 &	 57.4  (67)				\\
00093162009	&	2018-01-05	(58,123.00)	&	704.26	&	2.32	$\pm$	0.58	&	$1.55_{-0.31}^{+0.39} $	 &	 44.8  (42)				\\
00093162012	&	2018-01-19	(58,137.26)	&	1086.32	&	2.04	$\pm$	0.53	&	$1.26_{-0.28}^{+0.27} $	 &	 30.7  (47)				\\
00093162013	&	2018-01-21	(58,139.39)	&	1303.59	&	2.00	$\pm$	0.49	&	$1.33_{-0.24}^{+0.24} $	 &	 42.1  (56)				\\
00093162014	&	2018-01-26	(58,144.76)	&	1066.34	&	0.82	$\pm$	0.24	&	$2.02_{-0.40}^{+0.39} $	 &	 39.7  (33)				\\
00093162015	&	2018-02-02	(58,151.81)	&	973.94	&	2.35	$\pm$	0.73	&	$1.03_{-0.34}^{+0.34} $	 &	 31.2  (41)				\\
00094003001	&	2018-06-20	(58,289.90)	&	1133.77	&	1.00	$\pm$	0.37	&	$1.50_{-0.42}^{+0.56} $	 &	 24.9  (32)				\\
00094003002	&	2018-06-27	(58,296.33)	&	1463.41	&	1.18	$\pm$	0.38	&	$1.30_{-0.35}^{+0.35} $	 &	 34.0  (33)				\\
00094003004	&	2018-07-11	(58,310.61)	&	978.94	&	1.16	$\pm$	0.85	&	$1.19_{-0.61}^{+0.57} $	 &	 30.0  (26)				\\
00094003005	&	2018-07-18	(58,317.39)	&	1136.27	&	1.47	$\pm$	0.45	&	$1.21_{-0.35}^{+0.36} $	 &	 36.9  (37)				\\
00094003006	&	2018-07-25	(58,324.23)	&	1068.84	&	0.97	$\pm$	0.35	&	$1.58_{-0.49}^{+0.50} $	 &	 34.4  (29)				\\
00094003007	&	2018-08-01	(58,331.13)	&	936.00	&	1.25	$\pm$	0.43	&	$1.61_{-0.36}^{+0.36} $	 &	 27.9  (32)				\\
00094003008	&	2018-08-08	(58,338.45)	&	1088.82	&	0.83	$\pm$	0.24	&	$2.11_{-0.36}^{+0.41} $	 &	 15.8  (26)				\\
\hline
\end{tabular}
\end{center}
\tablecomments{Columns from left to right: (1) observation ID. (2) observation date. (3) net exposure time. (4) 0.3--10~keV model flux and its $1\sigma$ uncertainty. (5) Best-fit photon index and its $1\sigma$ lower and upper uncertainties. (6) C-Statistics and degree of freedom.
The corresponding X-ray observations of the four epochs (Pre-flare, Flare, Post-flare~I, and Post-flare~III) in Figure~\ref{fig:LC_hour} and Figure~\ref{fig:SEDFitting} are marked in column (1) by superscripts $a$, $b$, $c$, and $d$, respectively.}
\label{Table1}
\end{table*}

\subsection{Gamma-ray Observations: Fermi-LAT}
Using the Fermi data server\footnote{https://fermi.gsfc.nasa.gov/ssc/data/}, the newest Pass 8 data of PKS 1502+106 from October 10, 2014 to October 10, 2018, are acquired. Following the standard procedure\footnote{https://fermi.gsfc.nasa.gov/ssc/data/analysis/}, we use the \emph{Fermi Science Tools v10r0p5} with the \emph{P8R2\_SOURCE\_V6} instrument response function to analyze the acquired data. The 0.1--300 GeV events (evclass=128 and evtype=3) are extracted within a $10^{\circ}$ region of interest (ROI) centered on the location of PKS 1502+106 (R.A.=226.104, decl.=10.494, J2000) with \emph{gtselect}. In order to eliminate the Earth's limb events, the recommended quality cuts, (DATA\_QUAL$==$1)$\&\&$(LAT\_CONFIG$==$1), and a zenith angle cut at $90^{\circ}$ are applied. To prepare for variability and spectral analyses at different time-scales, a binned likelihood analysis to the total acquired data is first performed with \emph{gtlike} to obtain an initial spectral model. In this analysis, the input model file describing ROI is created using the \mbox{\emph{Fermi}-LAT} third source catalog (3FGL; \citealt{2015ApJS..218...23A}), together with the latest isotropic background model, \emph{iso\_P8R2\_SOURCE\_V6\_v06} and the galactic diffuse emission model, \emph{gll\_iem\_v06}.\footnote{https://fermi.gsfc.nasa.gov/ssc/data/access/lat/BackgroundModels.html} The model file contains sources within ROI+$10^{\circ}$ from the target. For sources lying within 10$^{\circ}$ from the center of ROI, their photon indexes and normalized parameters are left free to vary during the model fitting. For sources lying within $10^{\circ}$--$20^{\circ}$ from the center of ROI, their parameters are kept fixed to the 3FGL catalog values. In addition, the normalization of diffuse background components is kept free during the model fitting. 

Unbinned likelihood analyses are applied to extract fluxes and spectra under different time-scale requirements. In unbinned likelihood analyses, the initial spectral model obtained in the above binned likelihood analysis is used as the input model file. Considering the need for the convergence of likelihood fitting under short time-scale, for sources with the maximum-likelihood test statistic (TS; \citealt{1996ApJ...461..396M}) less than 25, their photon indexes and normalized parameters are set to be fixed regardless of their location. The fixed/free settings of parameters for other sources are consistent with that used in the binned likelihood analysis. Unless otherwise stated, the $\gamma$-ray spectrum of PKS 1502+106 is considered as a simple power-law form throughout the analyses. 

\subsection{X-ray Observations: Swift-XRT and XMM-Newton}
Using the astronomical archival data retrieval service provided by HEASARC,\footnote{https://heasarc.gsfc.nasa.gov/docs/archive.html} we search for the available Swift observations of PKS 1502+106 from October 10, 2014 to October 10, 2018. Fifty Swift observations are retrieved. Among them, 11 observations pointed to PKS 1502+106, and 39 observations pointed to Mkn~841. Mkn~841 is a Seyfert type-1 galaxy located about 7$'$ SW of PKS 1502+106, which allows the observations of Mkn~841 to provide the serendipitous \mbox{X-ray} data for PKS 1502+106. In order to obtain reliable photometric results, among the 50 observations, we only analyze the XRT data of 45 observations with net exposure time greater than 0.5 ks. The 45 observations are all made in the photon counting mode. The XRT data of these observations are processed by the task \emph{xrtpipeline} (v0.13.4). The calibration files (CALDB v20180710) and the standard filtering and screening criteria are used in this process. 

We use the \emph{xselect} tool to extract source photometry and spectrum, and we use the \emph{xrtmkarf} task to produce ancillary response file for each observation. A circular region with a radius of $47''$ centered at the object is used as the extracted area of the source, while a \mbox{source-free} annular region with a radius of $70''$ slightly away from the object is chosen as the extracted area of background.

We use XSPEC v12.9 to fit the extracted \mbox{0.3--10~keV} spectrum of each observation. Because of the low number of events from the source, events are not grouped, and $C$ statistics \citep{1979ApJ...228..939C} is employed to determine the best fit in the fittings. A power-law model modified by Galactic absorption ($N_\mathrm{H}= 2.20 \times 10^{20}$~cm$^{-2}$) is used in the fittings. 
The XRT observation logs of PKS 1502+106 are summarized in Table~\ref{Table1}, and the 0.3--10 keV model fluxes\footnote{The energy detection range of the XRT is 0.2--10~keV. However, in most observations, only 0.5--8 keV events (photons) are detected (due to the reduction of the effective area at high energy and low energy ends), which means that the model is mainly constrained by the 0.5--8 keV events.}, photon indices, and their $1\sigma$ uncertainties obtained from the spectral fittings are also listed in Table~\ref{Table1}.

In the 45 observations, the count rate of PKS 1502+106 (see Figure~\ref{fig:LC_3day}) is lower than the typical lower limit of the count rate ($\sim 0.5$~counts~s$^{-1}$) that will produce \mbox{pile-up}, suggesting that pile-up is unlikely to occur in these observations.
To further confirm whether pile-up exists, we analyze each observation following the standard pile-up analysis process\footnote{http://www.swift.ac.uk/analysis/xrt/pileup.php}, and we do not find any pile-up effect in these 45 observations. This result excludes the possibility that the hard photon indices ($\sim 1.0-1.3$) in Table~\ref{Table1} are due to the pile-up effect.

XMM-Newton performed an observation in the Imaging mode for Mkn 841 on July 14, 2015. At this time, PKS 1502+106 happened to be in the brightest stage in the $\gamma$-ray band (see Section~4 below). This observation, therefore, provides an important serendipitous X-ray data with a net exposure time of up to $\sim13$~ks for PKS 1502+106. The European Photon Imaging Camera (EPIC) on board the XMM-Newton is composed of three co-aligned X-ray telescopes, which simultaneously observe a source by accumulating photons in the three CCD-based instruments: the twins MOS1 and MOS2 and the pn. Here we only consider the \mbox{EPIC-pn} data as it is most sensitive. Following "The XMM-Newton ABC Guide" (v4.7), we use the XMM-Newton Science Analysis System (SAS v16.1.0) to extract the photometry and spectrum of PKS 1502+106. We extract the \mbox{$0.3$--$10$~keV} source events from a circular region with a radius of $45''$ centered on the object, and we extract the background events from a \mbox{source-free} circular region, with a radius of $70''$, slightly away from the object. Using the \emph{epatplot} task, we exclude the presence of pile-up. The standard background count rate threshold (RATE$<=0.4$) for screening good time intervals and the standard event pattern and flag filtering criteria (FLAG $==0$ \&\& PATTERN $<=4$) are applied in the extraction process. The tasks \emph{rmfgen} and \emph{arfgen} are used to create redistribution matrix file and ancillary response file, respectively. The extracted X-ray spectrum is grouped into at least 25 counts per energy bin to ensure the validity of $\chi^{2}$ statistics. Because the X-ray spectrum that obtained by the XMM-Newton has more photons, we first attempt to use a power-law model with both intrinsic absorption and Galactic absorption to fit it. The fitting result shows that the intrinsic absorption of PKS 1502+106 is very weak (the $90\%$ upper limit of $N_{\mathrm{H}}$ is $0.9 \times 10^{20}$~cm$^{-2}$), so we finally consider the power-law model only modified by Galactic absorption to fit its X-ray spectrum. The observation log and the X-ray spectrum fitting result are listed in Table~\ref{Table2}.

\begin{table*}
\tablenum{2}
\begin{center}
\caption{Observation of XMM-Newton EPIC-pn for PKS~1502+106}
\footnotesize
\begin{tabular}{@{}cccccc@{}}
\hline\hline
Obs-ID						&	Date (MJD)		&	Net Exp. 			&	Flux					&	$\Gamma_{\mathrm{X}}$		&						$\chi^{2}_{\mathrm{red}}$/dof 				\\
							&					&	(ks)				&	($10^{-12}$~erg~cm$^{-2}$~s$^{-1}$)					&						&									\\
(1)	&	(2)	&	(3)	&	(4)	&	(5)	 &	 (6)			\\
\hline
0763790501	&	2015-07-14	(57,217.83)	&	13.7	&	$1.61\pm	0.04$	& $1.67_{-0.02}^{+0.02} $ &	 0.91  (159)				\\
\hline
\end{tabular}
\end{center}
\tablecomments{The same format as Table~\ref{Table1} except that column (6) is reduced chi-square and degree of freedom.
The \mbox{X-ray} data of the Post-flare~II epoch in Figure~\ref{fig:LC_hour} and Figure~\ref{fig:SEDFitting} correspond to this observation.}
\label{Table2}
\end{table*}

\subsection{UV/Optical Observations: Swift-UVOT and Steward Observatory}
Another instrument on the Swift, UVOT, provides ultraviolet/optical data. In the 50 Swift observations carried out between October 10, 2014 and October 10, 2018, except for those observations where PKS~1502+106 falls outside the field of view, the UVOT data of 36 observations are available. The Swift-UVOT has three \mbox{optical-band} filters (V, B, U) and three \mbox{UV-band} filters (W1, M2, W2). Of the 36 available observations, 12 observations obtain six band data, and the remaining 24 observations only obtain the W2-band data. For each observation, we sum multiple images in the same filter with the task \emph{uvotimsum} and then perform aperture photometry with \emph{uvotsource}. Source counts are extracted from a circle with $5''$ radius centered on the object, while background counts are extracted from a circle with $20''$ radius located in a source-free region close to the object. The observed magnitudes are corrected for Galactic extinction using the reddening coefficient of $E_{(B-V)} = 0.0275$ mag from \citet{2011ApJ...737..103S} and \citet{1999PASP..111...63F} reddening law with $R_{V}=3.1$. The corrected observed magnitudes are then converted into fluxes using the zero points \citep{2011AIPC.1358..373B}. 

As a part of the Fermi multiwavelength support programme, the Steward Optical Observatory of the University of Arizona perform \mbox{long-term} optical monitoring for PKS 1502+106 (\citealt{2009arXiv0912.3621S}). The optical R-band and V-band photometric data and polarimetric data (optical linear polarization degree and polarization angle) of PKS 1502+106 can be easily acquired from the public archival database\footnote{http://james.as.arizona.edu/\%7epsmith/Fermi/}. 153 R-band and V-band photometric data and 188 polarimetric data are obtained from the optical monitoring between October 10, 2014 and October 10, 2018. Since the comparison star (SDSS J150418.48+102757.6) of PKS 1502+106 is not calibrated, only the differential magnitudes of PKS 1502+106 relative to the comparison star are provided in the database. In order to recover the apparent magnitudes of PKS 1502+106, we calculate the apparent magnitudes of the comparison star by utilizing its SDSS photometric data.\footnote{SDSS~J150418.48+102757.6 is a star and its SDSS magnitudes in the u, g, r, and i bands are $17.932\pm0.011$~mag, $15.984\pm0.004$~mag, $15.225\pm0.004$~mag, and $14.947\pm0.004$~mag, respectively \citep[][]{2015ApJS..219...12A}. Following the recipe of \citet{2005AAS...20713308L}, its apparent magnitudes in the R and V bands can be converted from the SDSS magnitudes.}
The estimated apparent magnitudes of the comparison star in the R and V bands are $15.0$ mag and $15.5$ mag, respectively.
Based on the estimated apparent magnitudes, we obtain the apparent magnitudes of PKS 1502+106, while we consider a large error ($0.1$ mag) on the recovered magnitudes account for the uncertainties introduced in the apparent magnitude estimation of the comparison star. The recovered magnitudes of PKS 1502+106 are corrected for Galactic extinction and then converted into fluxes.

\subsection{Radio Observations: OVRO 40m Telescope}
PKS 1502+106 is observed in radio band by the Owens Valley Radio Observatory (OVRO) 40m Telescope as a part of Fermi multiwavelength support programme \citep{2011ApJS..194...29R}. Through the database\footnote{http://www.astro.caltech.edu/ovroblazars/index.php?page=home} provided by the OVRO, we obtain the radio data of PKS 1502+106 detected at 15~GHz from October 10, 2014 to October 10, 2018.
\begin{figure*}
\centering
\includegraphics[width=6.3in]{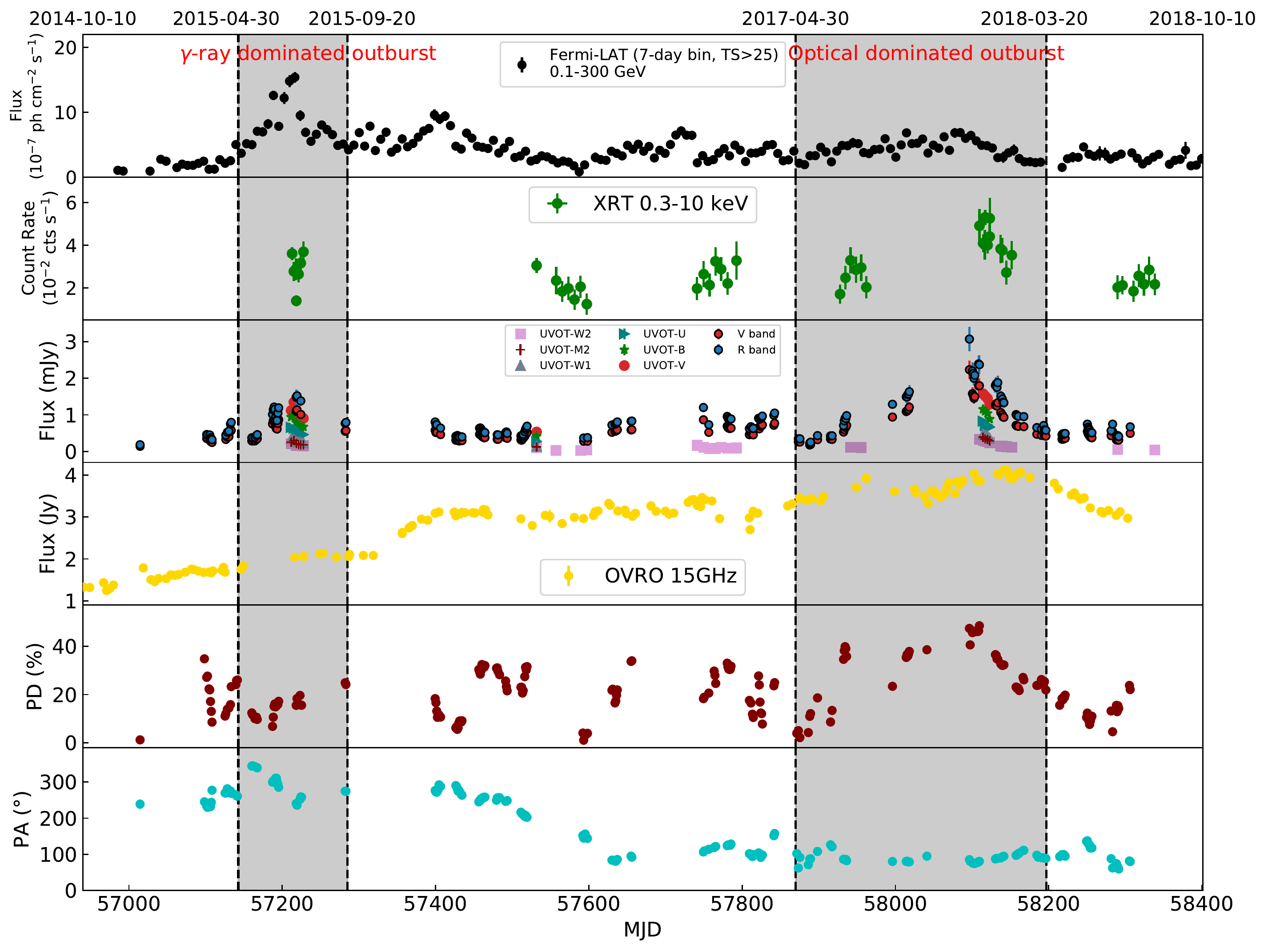}
\caption{Multiwavelength light curves of PKS 1502+106 from October 10, 2014 to October 10, 2018. From top to bottom, the light curves of $\gamma$-ray, X-ray, UV/optics as well as the changes of optical polarization degree (PD) and polarization angle (PA) are shown. There are two dramatic outbursts during this period. One is dominated by $\gamma$-ray emission and the other is dominated by optical emission. They are marked with grey shaded areas.}
\label{fig:LC}
\end{figure*}

\section{Global characteristics of multi-band variability}
In this section, we first summarize and analyze the global characteristics of the multiwavelength variability of PKS~1502+106 from October 2014 to October 2018. Then, two significant outbursts (see Figure~\ref{fig:LC}) in this period are investigated in detail in Sections 4 and 5 below.
\subsection{Multiwavelength light curves}
Figure~\ref{fig:LC} presents the multiwavelength light curves of PKS 1502+106 from October 2014 to October 2018. The first panel shows the 7-day binning light curve in the $\gamma$-ray band. 
Before April 30, 2015, PKS 1502+106 were in a quiescent stage with an average $\gamma$-ray flux of $1.1\times10^{-7}$~ph~cm$^{-2}$~s$^{-2}$. 

Between May 2015 and September 2015, a significant \mbox{$\gamma$-ray} outburst ($\gamma$-ray dominated outburst) occurred in PKS~1502+106. The peak flux in the $\gamma$-ray band was \mbox{$1.54\times10^{-6}$~ph~cm$^{-2}$~s$^{-2}$} at July 13, 2015 (for 7-day binning data). During this period, the \mbox{X-ray} and optical/ultraviolet fluxes showed associated changes. 
There was a gap period in the OVRO radio observations. Nevertheless, it can still be seen that the radio fluxes did not show significant changes, with a fractional variability\footnote{The fractional variability is usually used to reflect variability amplitude. It and its $1\sigma$ uncertainty are calculated using Eqs.~10 and 11 in \citet{2003MNRAS.345.1271V}.} of only ($6.8\pm0.2$)\%. During the outburst, the average optical linear polarization degree of PKS 1502+106 was ($15.1\pm0.5$)\%, and the fractional variability of the polarization degree was ($29.6\pm2.7$)\%; the polarization angle (PA) did not seem to have a significant swing of $>180^\circ$, and the maximum rotation amplitude was $\sim44.9^{\circ}$. Here, to solve the $\pm180^\circ$ ambiguity, following the recipe of \citet{2013ApJ...768...40L}, we added/subtracted $180^\circ$ each time that the subsequent PA value is $>90^\circ$ less/more than the preceding one. However, it should be noted that due to sparse polarization observations, the $\pm180^\circ$ ambiguity of PA may not be completely eliminated. There is still a possibility that the PA rotates an angle of $>90^\circ$ within a few days (see, e.g., \citealt{2008Natur.452..966M, 2013ApJ...768...40L, 2016A&A...590A..10K}), and such rotation events will be handled improperly by the current processing method, resulting in an underestimation of the variability of the PA.

During the period from September 20, 2015 to April 30, 2017, the $\gamma$-ray emission of PKS 1502+106 was in a plateau with average flux slightly higher than that of the quiescent stage (about four times higher); the $\gamma$-ray fluxes showed some minor fluctuations, with a fractional variability of ($38.0\pm2.7$)\%. In the X-ray and UV/optical bands, there were also only some minor fluctuations and no significant outburst. 
There were large random variations in the optical polarization degree, with a fractional variability of $(45.4\pm2.2)$\%.

The second significant outburst occurred between May 2017 and March 2018. It can be clearly seen that the optical emission of PKS 1502+106 showed a clear outburst profile, and the maximum flux in the R band reached $(3.1\pm0.3)$~mJy, about six times that of the quiescent stage. During this period, X-ray emission and polarization degree showed associated changes, but $\gamma$-ray fluxes did not show significant variations, 
with a fractional variability of ($28.1\pm1.6$)\%, so we called this outburst an optical dominated outburst.

\begin{figure}
\centering
\includegraphics[width=3.5in]{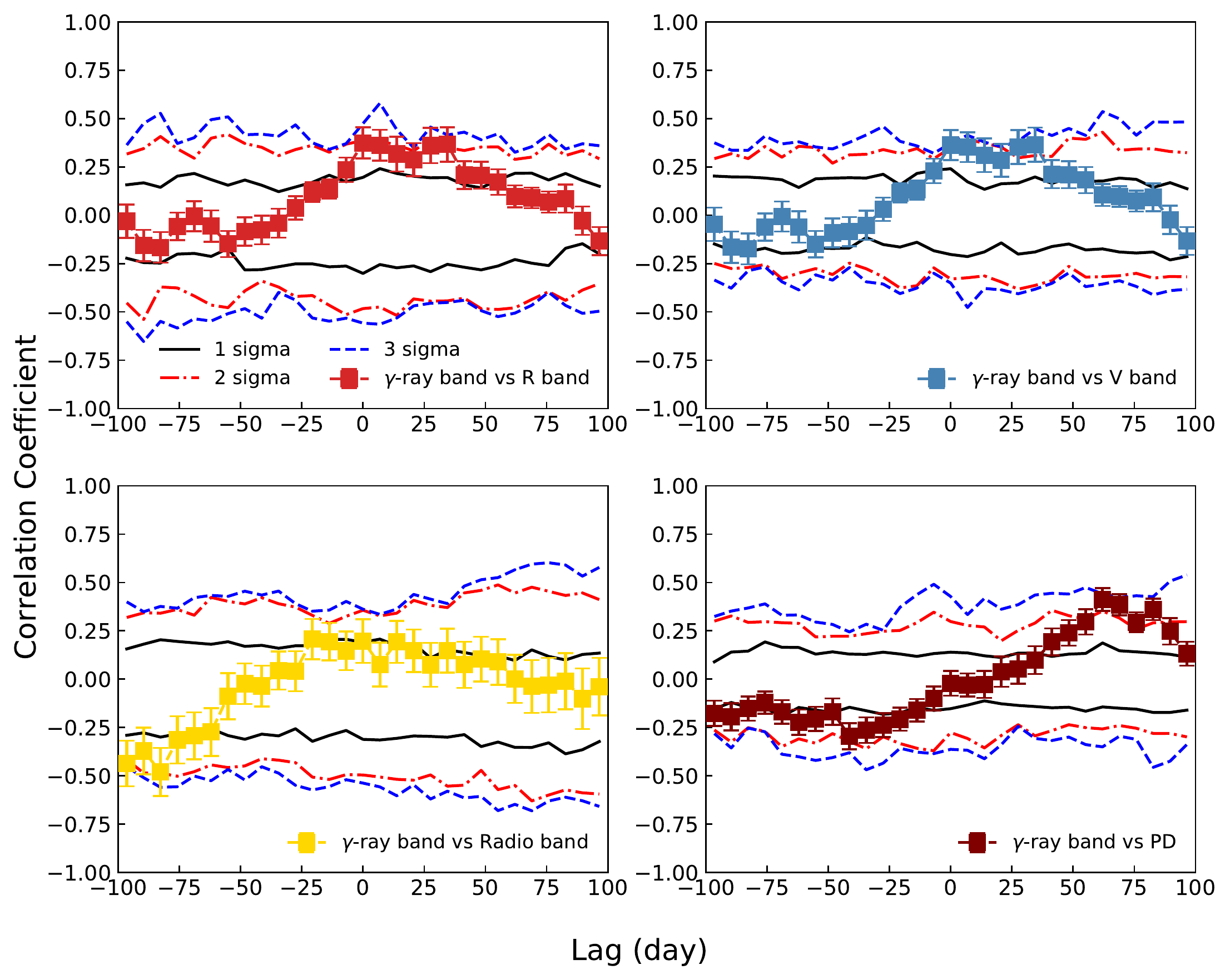}
\caption{DCF between the $\gamma$-ray emission and the emissions in the R, V, radio bands and the optical polarization degree (PD) over the whole observation period. The lines show the significance levels of DCF obtained by the MC simulations.}
\label{fig:DCF}
\end{figure}

\begin{figure*}
\centering
\includegraphics[width=5in]{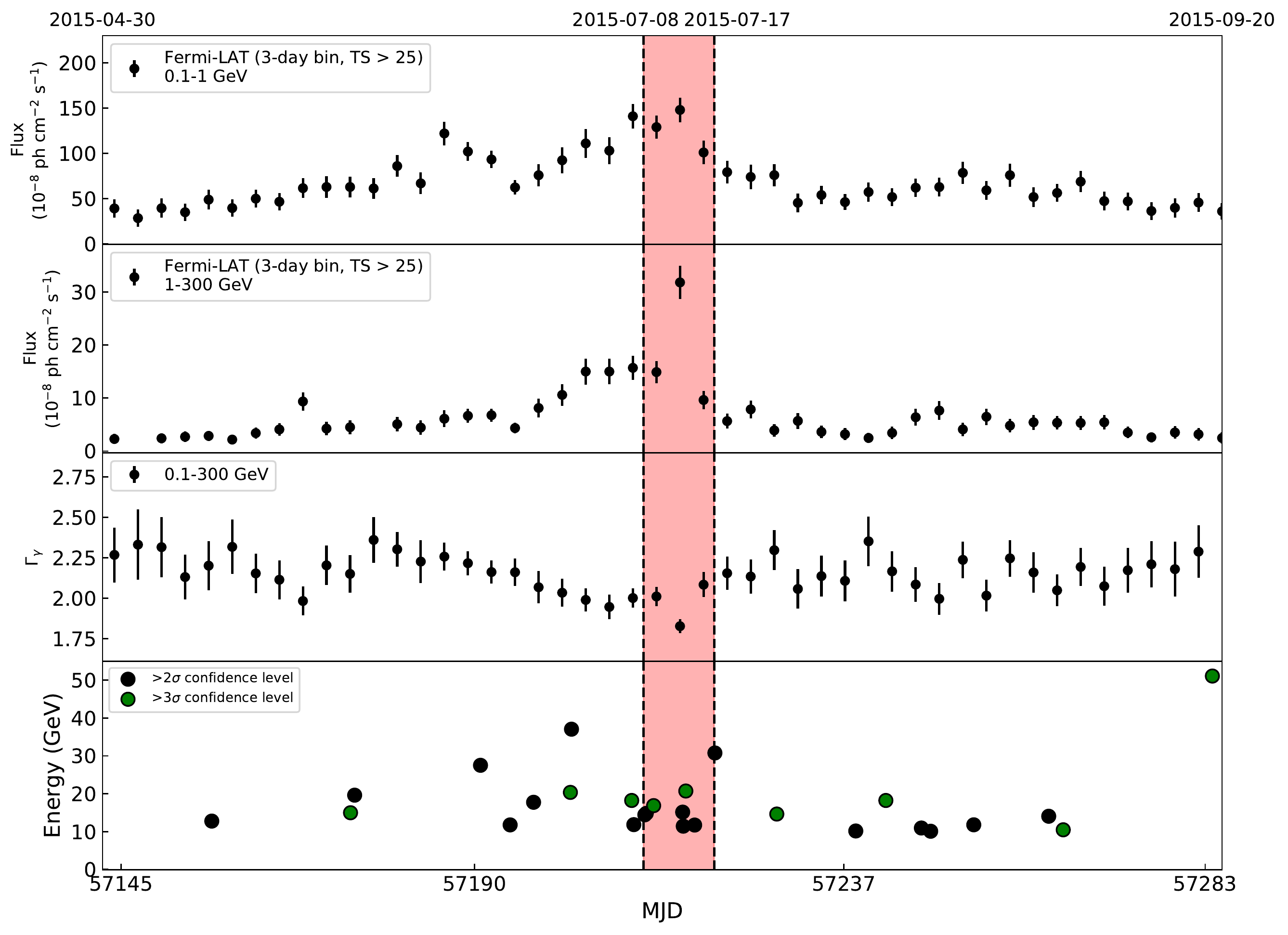}
\caption{Top and second panels: 3-day binning light curves in the 0.1--1~GeV and 1--300~GeV bands during the $\gamma$-ray dominated outburst in 2015. Third panel: the evolution of the \mbox{$\gamma$-ray} spectral index (0.1--300 GeV) with time. Bottom panel: the distribution of high energy events ($>$ 10 GeV) detected from PKS 1502+106. It can be clearly seen that a rapid GeV flare occurred in PKS 1502+106 from July 8 to July 17, 2015 (red shaded area). }
\label{fig:LC_3day}
\end{figure*}

\subsection{Multiwavelength correlations}
By investigating the correlation between different energy bands, we can obtain some information about the radiation mechanism and the location of the emission region in different bands \citep[see, e.g.,][]{2014MNRAS.445..428M, 2016MNRAS.456..171R, 2018MNRAS.480.5517L}. Therefore, we employ discrete correlation function (DCF, \citealt[][]{1988ApJ...333..646E}), which is one of the most common methods for studying cross-correlation between two time series with uneven spacing, to perform cross-correlation analyses on the multi-band data of PKS 1502+106. The significance of DCF is estimated by a Monte Carlo (MC) method introduced by \citet[][]{2014MNRAS.445..437M}. In MC simulations, we generate 1000 random light curves for each band with a simple power-law power spectral density (PSD) ($P(\nu) \propto \nu^{-\beta}$) following the algorithm described in \citet[][]{1995A&A...300..707T}; the spectral index $\beta$ for each band is determined by fitting the PSD of the actual observed light curve.

Figure~\ref{fig:DCF} shows the DCF between the $\gamma$-ray emission and the emissions in the R, V, radio bands and the optical polarization degree over the whole observation period, respectively. The 7-day binning interval is adopted in the analyses. The X-ray and UV bands are not included because they have fewer observation data. We do not find any significant \mbox{($>3\sigma$)} correlations between the $\gamma$-ray band and the optical bands (R and V bands). This may be due to the fact that PKS~1502+106 is in a steady state for most of the whole observation period. 
In addition, the thermal radiation from the accretion disk may also result in the weakening of the correlation between the optical emission and non-thermal \mbox{$\gamma$-ray} emission.
\citet[][]{2014MNRAS.445..428M} performed a cross-correlation analysis for the $\gamma$-ray data and 15 GHz radio data of PKS 1502+106 observed between 2008 to 2011 (MJD~\mbox{54600--55800}). They found a $2.25\sigma$ \mbox{confidence-level} positive correlation between the leading $\gamma$-ray emission and the lagging radio emission with a time delay of $(40\pm13)$ days. However, our results show that there is no significant ($>2\sigma$) correlation between the $\gamma$-ray emission and 15~GHz radio emission for the data observed from 2014 to 2018. This result implies that even for the same source, the correlation between different bands will behave differently at different periods \citep[][]{2019Galax...7...20B}. 
During \mbox{2008--2011}, 
the delay of radio emission and its associated changes with $\gamma$-ray emission could be explained by a moving disturbance propagating along the jet from inner region to downstream extended radio emission region \citep[][]{2014MNRAS.445..428M}. During \mbox{2014--2018}, the increase of radio fluxes have no obvious correlation with the $\gamma$-ray or other bands (see Figure~\ref{fig:LC}). At this time, the variability behavior of the radio emission may only be related to the local variation of downstream extended radio emission region. 
Recent studies have found that in some blazars, their $\gamma$-ray emission is associated with their optical polarization degree and the rotation of PA \citep[see, e.g., ][]{2016ApJ...833...77I, 2018MNRAS.474.1296B}. Since the ambiguity of the PA can not be completely eliminated (see Section 3.1 above), here we only analyze the cross-correlation between the polarization degree and the $\gamma$-ray emission. We do not find any significant ($>3\sigma$) correlation between the two physical quantities. 

\section{$\gamma$-ray dominated outburst in 2015}
During the period from May to September 2015, the Fermi-LAT detected a highest-level renewed $\gamma$-ray activity in PKS 1502+106 after the $\gamma$-ray activities in \mbox{2008--2009} (see ATel \#7592, ATel \#7801 as well as Figure~\ref{fig:LC}). The 3-day binning light curves in the 0.1--1~GeV and 1--300~GeV band during this period are shown in Figure~\ref{fig:LC_3day}. It can be clearly seen that a rapid GeV flare occurred in PKS 1502+106 from July 8 to July 17, 2015. 
The highest flux in the \mbox{1--300} GeV band is about six times higher than that in the average state, reaching ($3.2\pm0.3)\times10^{-7}$~ph~cm$^{-2}$~s$^{-1}$ (for 3-day~binning data). Figure~\ref{fig:LC_3day} also presents the evolution of the \mbox{$\gamma$-ray} spectral index (0.1--300 GeV) with time. Similar to the GeV flares of other blazars (e.g., 3C 454.3 \citep[][]{2010ApJ...721.1383A}, OT 081 \citep[][]{2018MNRAS.480.2324K}, CTA 102 \citep[][]{2018ApJ...863..114G}), the $\gamma$-ray spectrum hardens when the source becomes brighter. At the peak of the flare, the $\gamma$-ray spectral index reached the hardest $\Gamma_{\gamma}=1.82\pm0.04$, which is rarely seen in FSRQs. The distribution of high energy events ($>$ 10 GeV) detected from PKS 1502+106 is also presented in Figure~\ref{fig:LC_3day}. These high energy events are extracted using the \emph{gtsrcprob} tool within the $0.5^{\circ}$ ROI. Most of the high-energy photons were observed during the flare, and their energy mainly ranges from 10 to 20 GeV. Below we will combine the multi-band data to analyze this fast GeV flare to explore the possible origin mechanism of the flare and the physical properties of flaring region.

\begin{figure*}
\centering
\includegraphics[width=5in]{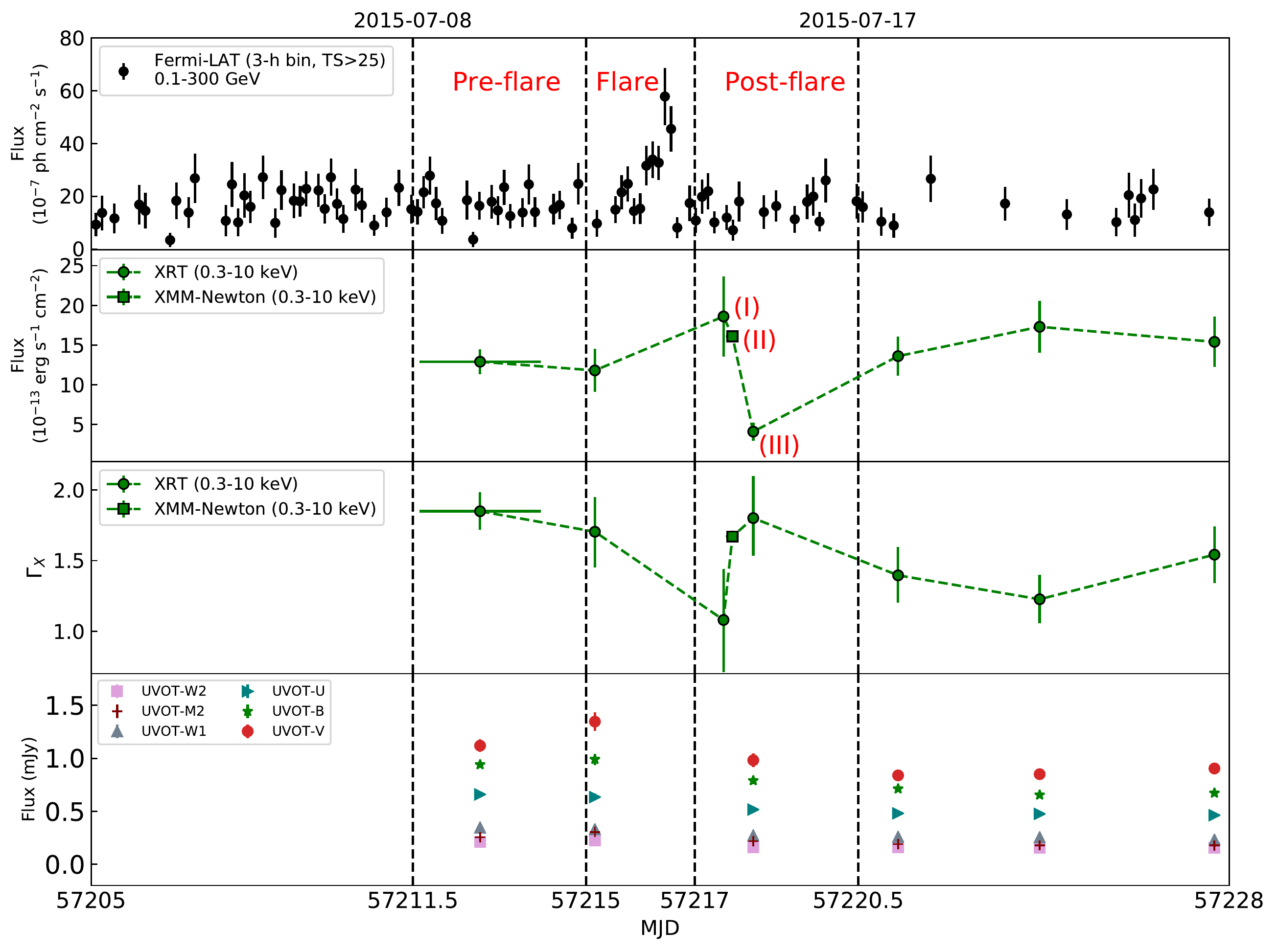}
\caption{Top panel: 3-h binning $\gamma$-ray light curve during the $\gamma$-ray flare in 2015. Second panel: X-ray light curve. Third panel: optical/UV light curves. Bottom panel: the changes of the X-ray photon index (0.3--10 keV) with time. As shown in the figure, we divide the flare into three phases (Pre-flare, Flare, and Post-flare). In the Post-flare phase, three independent X-ray observations are marked as (I), (II), and (III), respectively.}
\label{fig:LC_hour}
\end{figure*}

\begin{figure}
\hspace{-0.5cm}
\includegraphics[width=3.5in]{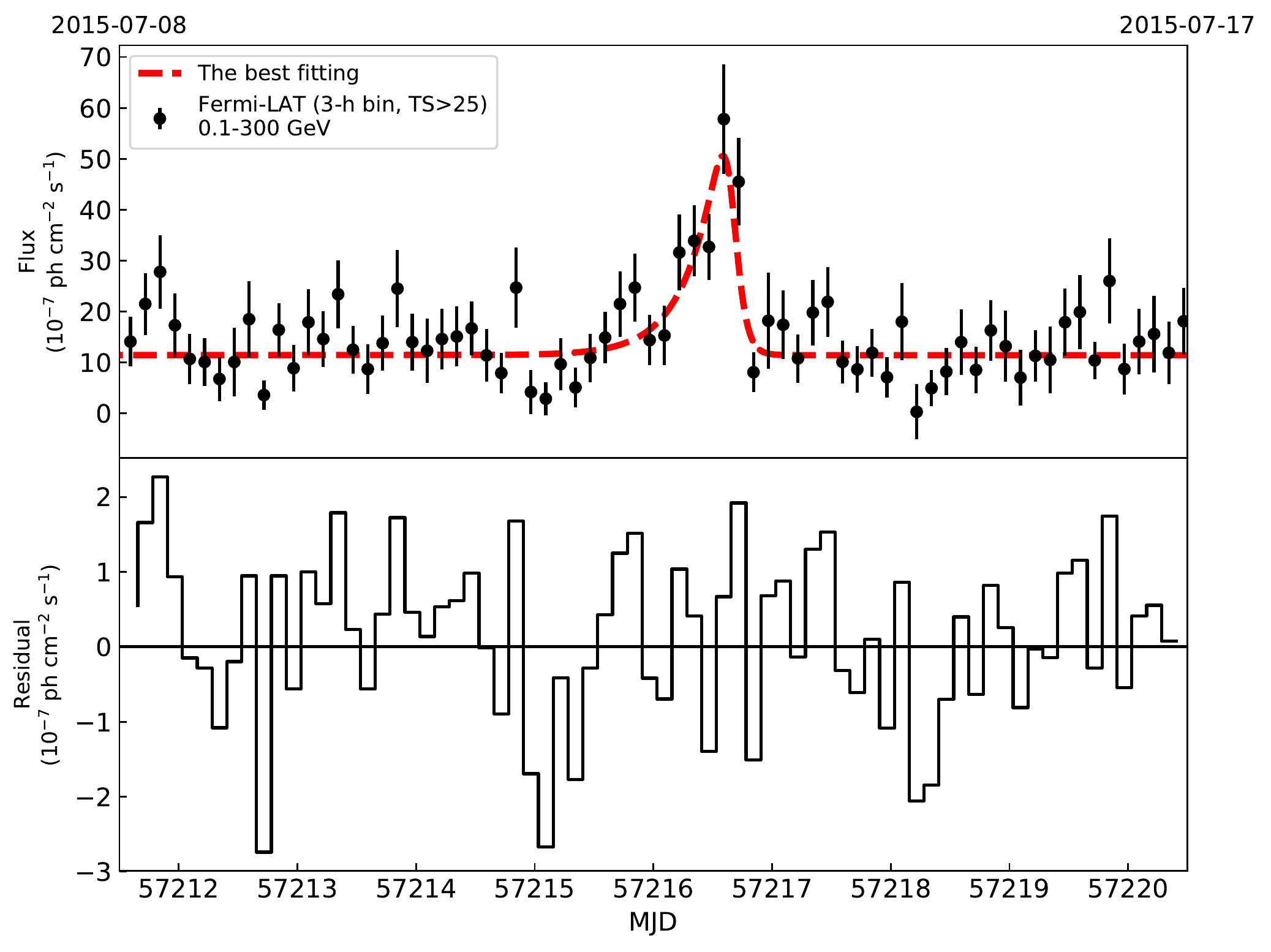}
\caption{Time-profile fitting (top panel) and residual (bottom panel) of the $\gamma$-ray flare in 2015 with 3-h binning.}
\label{fig:LC_fitting}
\end{figure}

\begin{figure*}
\centering
\includegraphics[width=7in]{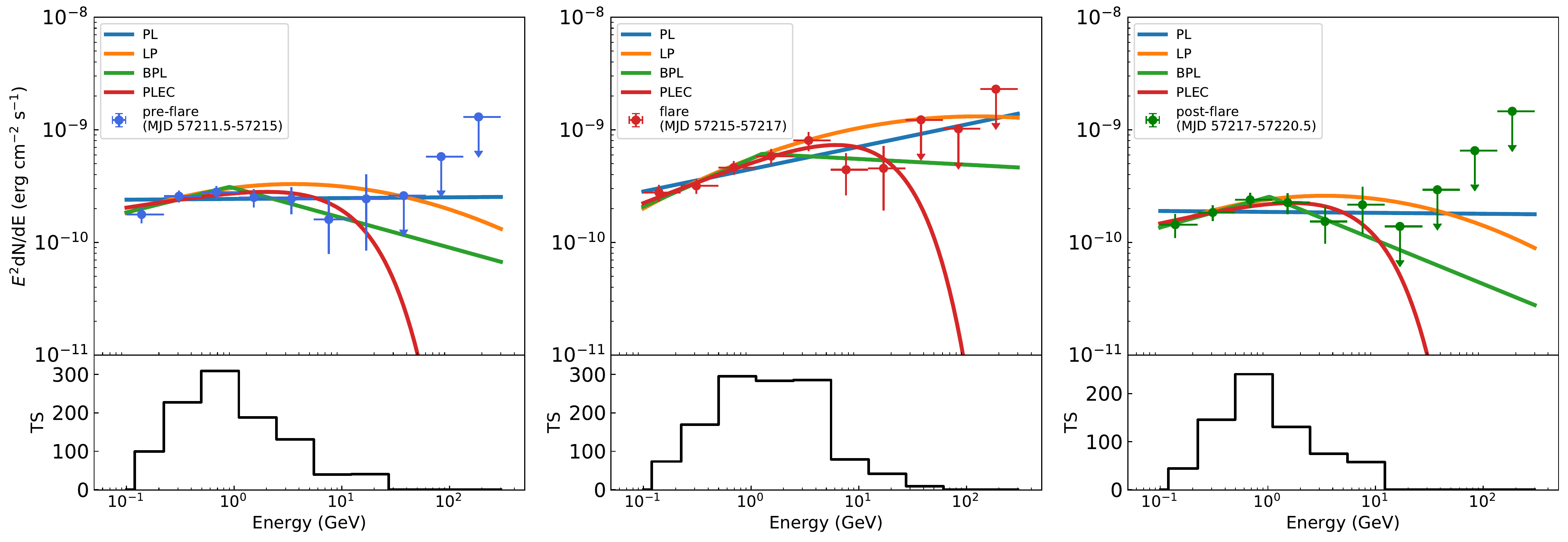}
\caption{$\gamma$-ray spectra and corresponding TS distributions at different phases of the flare in 2015. The blue, yellow, green and red lines show the unbinned likelihood spectral fitting results with PL, LP, BPL, and PLEC function forms, respectively. The corresponding fitting parameters are listed in Table~\ref{Table3}.}
\centering
\label{fig:GammaSED}
\end{figure*}

\subsection{The time profile of the flare}
High time-resolution light curve can eliminate the degeneracy caused by the superposition of multiple short time-scale variations and reveals more basic physical information. Therefore, on the premise of sufficient photon statistics, we obtain a 3-h binning $\gamma$-ray light curve during the flare (see Figure~\ref{fig:LC_hour}). Simultaneous optical/ultraviolet/X-ray observations and the changes of the X-ray photon index (0.3--10~keV) with time are also shown in this figure. The \mbox{$\gamma$-ray} flare is very fast, and there seems to be a related change in the X-ray band after the $\gamma$-ray flare. We divide the light curve into three phases (Pre-flare, Flare, and Post-flare) as shown in Figure~\ref{fig:LC_hour}. 

To quantitatively determine the rising and decaying time-scales of the flare, we fit the time profile of the flare with the following double exponential form function \citep[][]{2010ApJ...722..520A}:
\begin{equation}
\label{eq1}
F(t) = F_{c}+F_{0}\times(e^{\frac{t_{0}-t}{T_{r}}} + e^{\frac{t-t_{0}}{T_{d}}})^{-1},
\end{equation}
where $F_{c}$ is the quiescent flux; $F_{0}$ is the flux at time $t_{0}$, representing the amplitude of flare; $T_{r}$ and $T_{d}$ determine the rising and decaying time-scales of flare, respectively. The fitting result is shown in Figure~\ref{fig:LC_fitting}. The rising and decaying time-scales of the best fitting and their $1\sigma$ uncertainties are $T_{r} = (6.21\pm1.7)$~hr and $T_{d} = (1.46\pm0.5)$~hr, respectively. 
The flare possesses an obviously asymmetric shape (a relatively slow rise followed by a very rapid decay), with the symmetry parameter of $\xi=(T_{d}-T_{r})/(T_{d}+T_{r})=-0.61$. This implies that the injected/accelerated high-energy electrons are quickly cooled or escaped.

Based on the \mbox{$\gamma$-$\gamma$} opacity argument, the minimum Doppler factor can be estimated numerically from the maximum photon energy $E_{\max}$ detected during the flare \citep[][]{1995MNRAS.273..583D, 2010ApJ...717L.118P}. Assuming that the \mbox{$\gamma$-$\gamma$} interaction optical depth at $E_{\max}$ is one, the minimum Doppler factor can be calculated as
\begin{equation}
\label{eq2}
\delta_{\min} = (\frac{\sigma_{\mathrm{T}}d_{l}^{2}(1+z)^{2}f_{\epsilon}E_{\max}}{4t_{var}^{obs}m_{e}c^{4}})^{1/6},
\end{equation}
where $\sigma_{\mathrm{T}}$ and $d_{l}$ are the Thomson scattering \mbox{cross-section} and the luminosity distance, respectively. $t_{var}^{obs}$ is the shortest observed doubling/halving time-scale, and it is approximately equal to $\ln{2} \times T_{d} $\citep[see, e.g,][]{2018ApJ...863..114G}. During the flare, the maximum photon energy $E_{\max}$ is detected to be $\sim20$~GeV, and the contemporaneous X-ray flux (i.e., $f_{\epsilon}$) is $\sim1.5\times10^{-12}$~erg~cm$^{-2}$~s$^{-1}$ (see Figure~\ref{fig:LC_hour}), so that $\delta_{\min}\approx37$. For blazars, the bulk Lorentz factor $\Gamma$ of radiating blob 
generally have $\Gamma\sim\delta$, so an upper limit on the viewing angle of the jet can also be estimated: $\theta \lesssim 1/\delta_{\min}=1.5^{\circ}$. \citet[][]{2016A&A...586A..60K} performed ultra-high angular resolution \mbox{mm-VLBI} observations for PKS 1502+106. The observations show that PKS 1502+106 possesses a parsec-scale jet with the estimated Doppler factors ranging from $\sim7$ to $\sim50$ along the jet. The viewing angle differs between the inner and outer jet, with the former at $\theta\sim3^{\circ}$ and the latter at $\theta\sim 1^{\circ}$. The parameters constrained by the doubling/halving time-scales are compatible with that estimated from the mm-VLBI observations.

Based on the doubling/halving time-scale, the upper limit on the physical size of the flaring region can be given by
\begin{equation}
\label{eq3}
R \le ct_{var}^{obs}\frac{\delta_{\min}}{1+z} = 1.4\times10^{15}~\mathrm{cm}.
\end{equation}
This result indicates that the emission region of this flare is very compact, similar to findings in some \mbox{low-redshift} FSRQs (e.g., PKS~1510-089 \citep{2010MNRAS.405L..94T, 2017ApJ...844...62P}, CTA~102 \citep{2018ApJ...854L..26S}). For a conical jet, if the emission is produced across the entire jet area, such a compact emission region implies that the flaring region is very close to the central engine; the upper limit on the distance of the flaring region from the central engine is 
$d \sim 2R/\theta \sim2c\delta_{\min}^{2}t_{var}^{obs}/(1+z)=0.03~\mathrm{pc}$ \citep[][]{2011ApJ...733L..26A}. 

\subsection{The evolution of $\gamma$-ray spectrum}
To quantify the curvature of $\gamma$-ray spectra at different phases of the flare and search for possible spectral evolution, we perform unbinned likelihood spectral fittings for the $\gamma$-ray spectra at the Pre-flare, Flare, and Post-flare phases with power law (PL), log parabola (LP), broken PL (BPL), and power law with an exponential cutoff (PLEC) function forms. These functions are commonly used to analyze the $\gamma$-ray spectra of blazars \citep[see, e.g.,][]{2010ApJ...721.1383A, 2011ApJ...733L..26A}. The fitting results of the $\gamma$-ray spectra at different phases are presented in Figure~\ref{fig:GammaSED}, and the corresponding fitting parameters are listed in Table~\ref{Table3}. As mentioned in \citet[][]{2011ApJ...733L..26A}, although the likelihood analysis can return the log-likelihood value along with the fitting parameters, the log-likelihood value does not provide an absolute \mbox{goodness-of-fit} evaluation. Therefore, the reduced $\chi^2$ value of each fitting is also calculated and listed in Table~\ref{Table3}. Following \citet[][]{2012ApJS..199...31N}, we use the $\mathrm{TS}_{\mathrm{curve}}$ value, which is calculated by $\mathrm{TS}_{\mathrm{curve}} = 2(\log \mathcal{L}(\mathrm{LP/BPL/PLEC}) - \log \mathcal{L}(\mathrm{PL}))$, to evaluate the significance of spectral curvature.

As shown in Figure~\ref{fig:GammaSED}, at the Flare phase, the $\gamma$-ray spectrum is rather hard, and a curvature/truncation feature emerges. The significance level of the curvature/break is $\mathrm{TS}_{\mathrm{curve}} = 10.26$ ($\sim 3.2 \sigma$) under the PLEC function form. In all cases, the fitting of the LP function is the worst than that of other functions, and the PLEC function is the best fitting form. There is no obvious evolution in the break/cutoff energy between the three phases and only a slight increase in the break/cutoff energy at the Flare phase. This feature is similar to that found in low-redshift FSRQs (e.g., 3C454.3 \citep{2011ApJ...733L..26A}, PKS 1510-089 \citep{2017ApJ...844...62P}). 

There are four scenarios that are usually used to explain the observed spectral curvature/break \citep[see, e.g.,][]{2001ApJ...561..111G, 2009ApJ...699..817A, 2010ApJ...717L.118P, 2010ApJ...721.1383A, 2014ApJ...794....8S}: 1) the "cooling break" resulting from the radiative losses of high-energy electrons, 2) the attenuation of high-energy photons by the extragalactic background light (EBL), 3) the transition of Compton scattering \mbox{cross-section} from the Thomson regime to the Klein-Nishina (K-N) regime, and 4) the \mbox{photon-photon} pair absorption associated with the {He~\sc ii}~Lyman~continuum (LyC) or H~LyC. For PKS 1502+106, although it has a relatively high redshift (z = 1.838), the EBL at this redshift will only affect the spectra at energies $\gtrsim 20$~GeV in the observation frame \citep[][]{2011MNRAS.410.2556D}. In addition, the typical break energy of spectrum produced by the \mbox{photon-photon} pair absorption associated with the {He~\sc ii}~LyC is~$\sim 1.8$~GeV in the observation frame, and that associated with the H~LyC is~$\sim 7$~GeV in the observation frame \citep{2014ApJ...794....8S}. The spectral break energies observed in the three phases are $\sim 1$~GeV (see BPL fittings in Table~\ref{Table3}), which is less than the typical break energies in the above two cases. This implies that the observed spectral curvature/break here is unlikely to be caused by the external \mbox{photon-photon} pair absorption. It can be found that the difference between the high-energy spectral index and the low-energy spectral index obtained by the BPL fitting in the three phases is $\sim0.5$ (see Table~\ref{Table3}), which is consistent with the expectation of the typical "cooling break" \citep[][]{2001ApJ...561..111G, 2009ApJ...699..817A}. It is suggested that the observed spectral curvature/break is most likely caused by the radiative losses of high-energy electrons. 

\begin{figure*}
\centering
\includegraphics[width=7in]{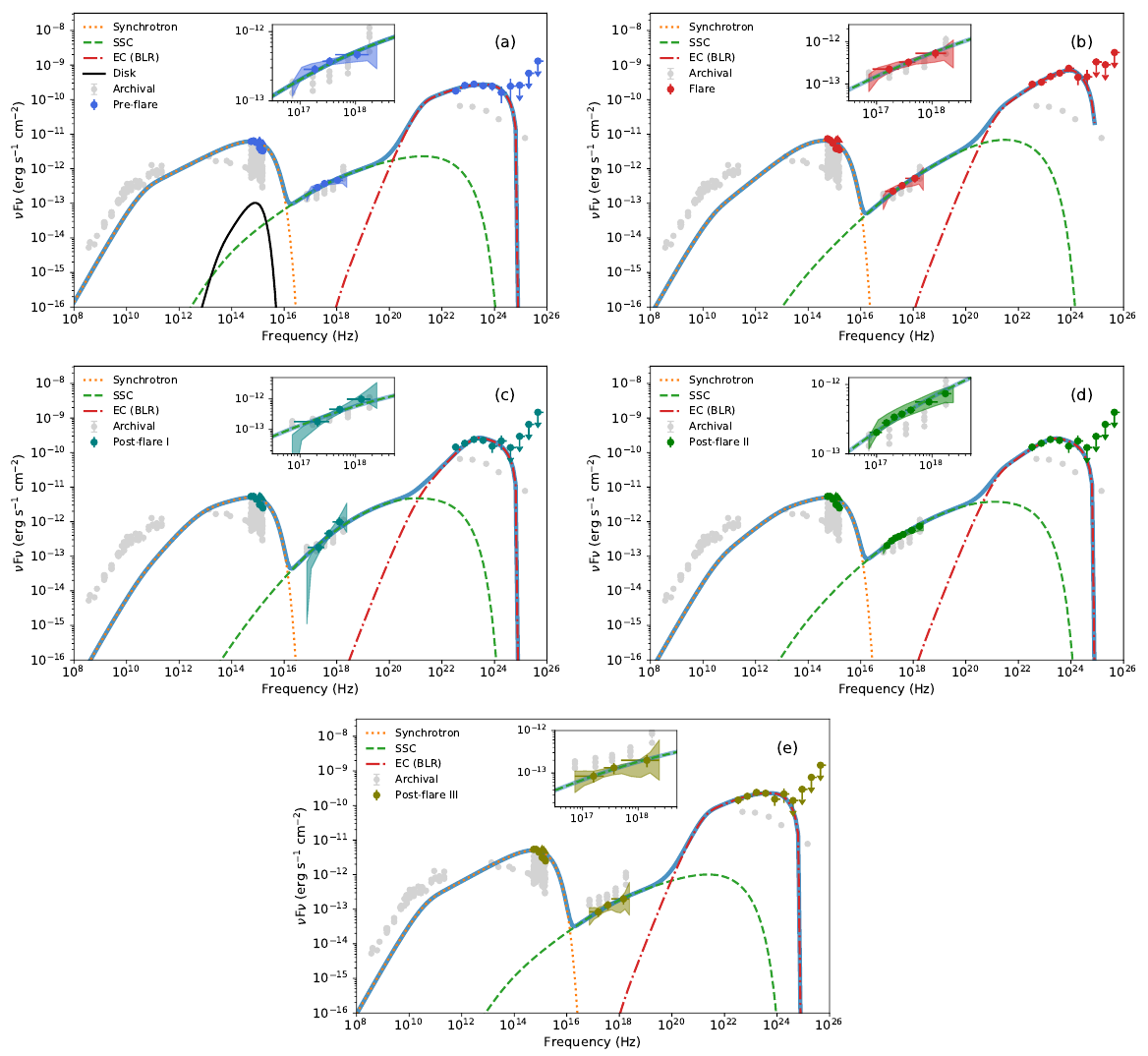}
\caption{Modelings of the multiwavelength SEDs of PKS 1502+106 at different epochs of the $\gamma$-ray flare in 2015. The corresponding optimal parameters of the multiwavelength SED modelings are listed in Table~\ref{Table4}. Here the SEDs of Post-flare~I, Post-flare~II and Post-flare~III correspond to three different X-ray observations at the Post-flare phase, respectively (see~Figure~\ref{fig:LC_hour}). An enlarged view in the \mbox{X-ray} band is shown in the inset of each panel.}
\label{fig:SEDFitting}
\end{figure*}

\subsection{The modeling and evolution of multi-band SED}
To further investigate the underlying trigger mechanism of the rapid $\gamma$-ray flare, we model the simultaneous multi-band SEDs at different epochs. The \mbox{multi-band} SEDs at different epochs are shown in Figure~\ref{fig:SEDFitting}. 
Here the SEDs of Post-flare~I, Post-flare~II and Post-flare~III correspond to three different X-ray observations at the Post-flare phase, respectively (see~Figure~\ref{fig:LC_hour}); they have the same data in the UV/optical and $\gamma$-ray bands. 
In each SED, the \mbox{X-ray} data points are the actual observed \mbox{X-ray} data after grouping, and each energy bin has no less than 15 photons. These observed \mbox{X-ray} data points are used to perform SED modeling, while a butterfly plot obtained from the X-ray spectral fitting is displayed in the \mbox{X-ray} band as a reference. 
The grey data points are archive data collected by the ASDC SED Builder Tool, an online service developed at the ASI Science Data Center \citep[][]{2011arXiv1103.0749S}.\footnote{The archive data in the $\gamma$-ray band are the integral data of \mbox{\emph{Fermi}-LAT} three-year observations (i.e. three-year binning data), so the source exhibits the $\gamma$-ray detection in the $> 50$~GeV band even if the source is in a low state. On the contrary, since the integral time of the $\gamma$-ray data obtained during the flare is very short (2-day bin), even if the source is in a high state, only the upper limits are obtained in the $> 50$~GeV band.}
These archive data are not involved in the SED modeling and are only for reference.
At high redshift, the Lyman-alpha forest produced by neutral hydrogen absorption shifts to the low-frequency band. For a neutral hydrogen cloud with $z\sim1.8$, their Lyman-alpha absorption starts at about 343~nm in the observation frame, which means that the UV photometric data of the Swift-UVOT may suffer from the influence of Lyman-alpha absorption (see, e.g., \citealt[][]{2010A&A...509A..69B, 2012A&A...538A..26R}). Therefore, the UV data in here are considered as the lower limits.

Below we model the SEDs in the framework of a conventional one-zone homogeneous leptonic model (see, e.g., \citealt{2009MNRAS.397..985G, 2016MNRAS.459.3175Y, 2017MNRAS.464..599D}, for detailed model description). It should be noted that lepto-hadronic models can also reproduce the \mbox{multi-band} emission of blazars \cite[e.g.,][]{2012arXiv1205.0539B,2013ApJ...768...54B}. In the lepto-hadronic models, the emission of X-ray to high-energy $\gamma$-ray is generated by proton-photon interactions \citep[][]{1992A&A...253L..21M} or relativistic proton synchrotron radiation \citep[][]{2000NewA....5..377A}. Typically, however, the lepto-hadronic models are difficult to account for fast $\gamma$-ray flare owing to the long cooling time scales of protons \citep[see][]{2012arXiv1205.0539B}. In addition, for FSRQs, the lepto-hadronic models usually require an extremely high (super-Eddington) proton power (see, e.g., \citealt{2009ApJ...704...38S}; \citealt{2015MNRAS.450L..21Z}). 
In view of these, we will not explore the lepto-hadronic scenario in detail.
In the leptonic model, we assume that multi-band emission is produced by relativistic electrons in a spherical blob with radius $R_\mathrm{size}$ and filled with a uniform magnetic field $B$. The radiating blob moves in the direction close to the line of sight, and the bulk Lorentz factor has $\Gamma \sim \delta$. We assume that the electron energy distribution is a commonly used broken power-law form as follows:
\begin{equation}
\label{eq4}
N(\gamma) = N_{0}n(\gamma) = \left\{\begin{array}{rcl}
N_{0}\gamma^{-p_{1}}&      & \gamma_{\min}  < \gamma     \le      \gamma_{\mathrm{br}}\\
N_{0}\gamma^{-p_{2}}\gamma_{\mathrm{br}}^{p_{2}-p_{1}}&      & \gamma_{\mathrm{br}}    < \gamma  \le     \gamma_{\max},\\
\end{array} 
\right.
\end{equation}
where $N_{0}$ is the number of emitting particles per unit volume; $\gamma_{\min}$, $\gamma_{\mathrm{br}}$, and $\gamma_{\max}$ are the electron Lorentz factors for minimum, break, and maximum, respectively; $p_{1}$ and $p_{2}$ are indices of the power law below and above the break energy.
In the leptonic model, the observed high-energy emission is produced by the inverse Compton (IC) scattering of relativistic electrons. The seed photons for IC process could be from the local synchrotron emission (i.e., synchrotron self-Compton, SSC) or from external fields (EC), such as broad-line region (BLR) and dusty torus (DT). As shown in Figure~\ref{fig:SEDFitting}, in most cases, the shapes of the energy spectra in the \mbox{X-ray} band and the $\gamma$-ray band are similar. Such SED can not be reproduced well with SSC model only and needs the involvement of an EC component. The type of soft photons in the EC process depends on the location of the emission region. For PKS~1502+104, its monochromatic luminosity at 1350~\AA~is $L_{1350} = 7.8 \times 10^{46}$~erg~s$^{-1}$~cm$^{-2}$ \citep{2012ApJ...748...49S}. According to the \mbox{C~IV~radius--$L_{1350}$ relation} obtained from the reverberation mapping of luminous quasars at high redshift \citep{2018ApJ...865...56L}, the BLR radius of PKS~1502+106 is estimated to be $R_{\mathrm{BLR}} \approx 0.11$~pc. This size is larger than the upper limit on the distance of the emission region from the central engine (see Section~4.1), suggesting that the photons coming from the BLR are dominant in the EC process. The central black hole mass ($M_{\mathrm{BH}} = 7.9 \times 10^{8}~M_{\odot}$) and the BLR luminosity ($L_{\mathrm{BLR}} = 1.47 \times 10^{45}$~erg~s$^{-1}$~cm$^{-2}$) of PKS 1502+106 are estimated by \citet[][]{2014MNRAS.441.3375X} based on the C~IV emission line. Following \citet[][]{2010MNRAS.402..497G}, the luminosity of accretion disk can be estimated as $L_{\mathrm{disk}} \approx 10\times L_{\mathrm{BLR}} = 1.47 \times 10^{46}$~erg~s$^{-1}$~cm$^{-2}$. Based on these values, the radiation spectrum from a standard thin disk is calculated and shown in Figure~\ref{fig:SEDFitting}~(a) \cite[]{1973A&A....24..337S}. It can be seen that even in the UV/optical band, the contribution of the emission from the disk is much smaller than the observed SED, indicating the observed SEDs are dominated by non-thermal radiation. For these reasons, we finally adopt SSC+EC (target photons coming from BLR) model to fit observed SEDs. The BLR photon density in the comoving/jet frame can be calculated as $U_{\mathrm{BLR}} = \frac{\Gamma^{2}L_{\mathrm{BLR}}}{4\pi cR_{\mathrm{BLR}}^{2}} \approx 0.03\delta^{2}~\mathrm{erg~cm}^{-3}$, and a effective BLR temperature $T_{\mathrm{BLR}} \sim 5 \times 10^{4}$~K is adopted in the model.
In addition, based on the EBL model in \citet[][]{2011MNRAS.410.2556D}, we consider the effect of EBL absorption in the GeV band in the model calculation.

There are nine free parameters in the model. Six of them specify the electron energy distribution ($N_{0}$, $\gamma_{\min}$, $\gamma_{\max}$, $\gamma_{\mathrm{br}}$, $p_{1}$, and $p_{2}$), and other three describe the properties of the emission region ($\delta$, $R_\mathrm{size}$ and $B$). 
The Doppler factor, the minimum and maximum energies of electrons are difficult to be constrained well by the current SED data. Therefore, in the model, we adopt the minimum Doppler factor of $\delta_{\min}  = 37$ and an appropriate minimum electron energy of $\gamma_{\min} = 20$, which is constrained by the archival radio data and the observed \mbox{X-ray} data in some degree.
In addition, based on the constraint of SED shape, the values of $\gamma_{\max} = 5\times10^{3}$ and $p_{2} = 2.5$ are adopted in the model to further reduce the number of free parameters. The final model free parameters are reduced to five. We employ the Levenberg-Marquardt algorithm, which is provided by the Python \emph{lmfit} package\footnote{https://lmfit.github.io/lmfit-py/}, to obtain the best-fitting parameters of the model for each observed SED. The best modeled SEDs are displayed in Figure~\ref{fig:SEDFitting}, and the corresponding model parameters are listed in Table~\ref{Table4}. 
The powers carried by the jet in the form of radiation ($P_{r}$), magnetic field ($P_{B}$), electrons ($P_{e}$) and cold protons ($P_{p}$, assuming one proton per emitting electron) are also given in Table~\ref{Table4}. They are calculated as
\begin{equation}
\label{eq5}
P_{i} = \pi R_{\mathrm{size}}^{2}\Gamma^{2}cU_{i} \approx \pi R_{\mathrm{size}}^{2}\delta^{2}cU_{i} ,
\end{equation}
where $U_{i}$ is the energy density of the $i$-th component \citep[see, e.g.,][]{2008MNRAS.385..283C, 2017MNRAS.464..599D, 2019ApJ...871...81X}.

The results show that the radiative power of the jet ($P_{r}$) is higher than the kinetic power carried in relativistic electrons ($P_{e}$) and the power in Poynting flux ($P_{B}$), which is consistent with the results of typical \mbox{high-power} FSRQs \citep[][]{2008MNRAS.385..283C}. In the framework of the leptonic model, this result implies that the jet may be mainly loaded by hadrons, namely dynamically dominated by the bulk motion of cold protons, as both leptons and Poynting flux do not provide sufficient power to account for the observed emission (see \citealt{2008MNRAS.385..283C, 2010MNRAS.402..497G} for detailed discussions). In addition, the power in Poynting flux is roughly comparable to the kinetic power carried in relativistic electrons, and the magnetization parameter ($P_{B}/P_{e}$) ranges from 0.2 to 1.5, which indicates that the magnetic field energy and electronic kinetic energy of the jet are basically in equipartition.

\begin{figure}
\centering
\includegraphics[width=3in]{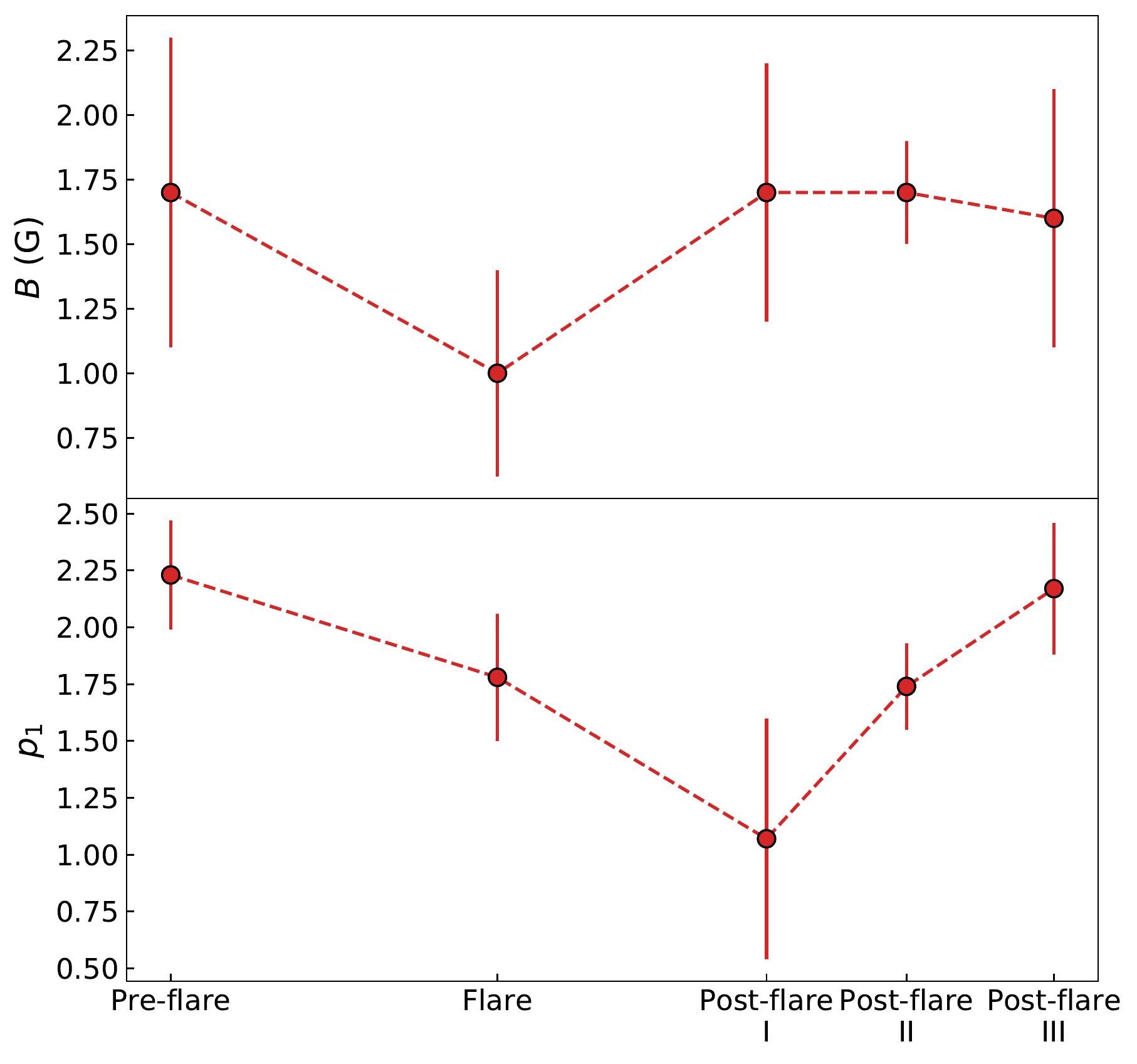}
\caption{The changes of the magnetic field intensity $B$ and the electron spectrum index $p_{1}$ at different epochs of the \mbox{$\gamma$-ray} flare in 2015.}
\label{fig:evolution}
\end{figure}

The sizes of emission region obtained by the SED fittings range $(3-8)\times 10^{15}$~cm, which is slightly larger than that constrained by the doubling/halving time-scale. Considering the uncertainties in the time-scale measurement and the SED fittings, the sizes obtained by the two methods are basically consistent. 
The values of the magnetic field intensity $B$ and the electronic spectral index $p_{1}$ at different epochs are shown in Figure~\ref{fig:evolution}. $p_{1}$ shows a soft-hard-soft change. It should be pointed out that changing the values of $\delta$, $p_{2}$, $\gamma_{\min}$, and $\gamma_{\max}$ will not substantially affect the fitting results of $p_{1}$.
In the Flare phase, the magnetic field intensity seems to decrease slightly, and then a harder electron spectrum with $p_{1} = 1.07\pm0.53$ appears in the Post-flare~I epoch.
Such a hard electronic energy spectrum usually can not be produced in the standard diffusive shock acceleration models, which generally forms an electronic energy spectrum with the electronic spectral index close to 2.0 \citep[see, e.g.,][]{1983RPPh...46..973D, 1998PhRvL..80.3911B, 2007Ap&SS.309..119R}. As thus, the fast $\gamma$-ray flare is probably not caused by shocks.
Note that an electronic spectral index of 1.5 is generally considered as the lower limit still in agreement with the shock acceleration scenarios \citep[e.g.,][]{2006Natur.440.1018A}. Considering the uncertainty of $p_{1}$, the shock acceleration scenarios can not be eliminated completely in here. However, magnetic reconnection may be a more promising trigger for this fast gamma-ray flare.
Three-dimensional numerical simulations show that the relativistic turbulence in AGN jet may trigger magnetic reconnection, which could further drive strong stochastic acceleration and form a hard electronic energy spectrum with an electronic spectral index of $\sim1$ \citep[see][]{2014PhRvL.113o5005G, 2015ApJ...806..167G}.

\begin{table*}
\tablenum{3}
\centering
\caption{Summary of the Likelihood Fitting Results of $\gamma$-ray Spectra at Different Phases of the Flare in 2015.}
\scriptsize
\begin{tabular}{@{}cccccccc@{}}
\hline\hline
\multicolumn{7}{c}{\bf{Power Law (PL)}}\\
\hline
Phase						&	$F_{0.1-300~\mathrm {GeV}}$ 					&	$\Gamma$				&						&			&		$-\log(\mathrm{Likelihood})$				&$\chi^{2}_{\mathrm{red}}$&	 	\\
							&	      	($10^{-6}$ ph cm$^{-2}$ s$^{-1}$)					&					&						&								&						&&	\\
\hline
Pre-flare						&	1.51	$\pm$	0.11	&	1.99	 $\pm$	0.05			&						&						&				8761.88		&1.36&				\\
Flare						&	2.21	$\pm$	0.16	&	1.80	 $\pm$	0.04			&						&						&				6289.04		&1.56&				\\
Post-flare 					&	1.17	$\pm$	0.11	&     2.01 $\pm$	0.07			&						&						&				7738.03		&1.19&				\\
\hline
\multicolumn{7}{c}{\bf{Log Parabola (LP)}}\\
\hline
Phase						&	$F_{0.1-300~\mathrm {GeV}}$					&	$\alpha$				&		$\beta$				&			&		$-\log(\mathrm{Likelihood})$				&$\chi^{2}_{\mathrm{red}}$&	$\mathrm{TS}_{\mathrm{curve}}$ 	\\
							&	      	($10^{-6}$ ph cm$^{-2}$ s$^{-1}$)					&					&						&								&						&&	\\
\hline
Pre-flare						&	1.38	$\pm$	0.12	&	1.78	 $\pm$	0.09			&	0.11	$\pm$	0.04				&					&	8757.59					&2.18&			8.58				\\
Flare						&	2.00	$\pm$	0.17	&	1.56	 $\pm$	0.11			&	0.08	$\pm$	0.03				&					&	6285.18					&4.32&			7.72				\\
Post-flare 					&	1.05	$\pm$	0.12	&     1.76 $\pm$	0.13			&	0.12	$\pm$	0.06				&					&	7735.03					&1.41&			6.00				\\
\hline
\multicolumn{7}{c}{\bf{PLExpCutoff (PLEC)}}\\
\hline
Phase						&	$F_{0.1-300~\mathrm {GeV}}$					&	$\Gamma_{\mathrm{PLEC}}$				&		$E_{\mathrm{cutoff}}$				&			&		$-\log(\mathrm{Likelihood})$				&$\chi^{2}_{\mathrm{red}}$&	$\mathrm{TS}_{\mathrm{curve}}$ 	\\
							&	      	($10^{-6}$ ph cm$^{-2}$ s$^{-1}$)					&					&		(GeV)				&								&&						&	\\
\hline
Pre-flare						&	1.43	$\pm$	0.12	&	1.84	 $\pm$	0.09			&	12.6  $\pm$	7.1		&							&			8758.44			&0.95&		6.88		\\
Flare						&	2.05	$\pm$	0.16	&	1.61  $\pm$	0.08			&	16.3  $\pm$	7.2		&							&			6283.91			&0.92&		10.26		\\
Post-flare 					&	1.08	$\pm$	0.12	&     1.78	 $\pm$	0.12			&	7.6 $\pm$ 4.3			&							&			7734.97			&0.80&		6.12		\\
\hline
\multicolumn{7}{c}{\bf{Broken PowerLaw (BPL)}}\\
\hline
Phase						&	$F_{0.1-300~\mathrm {GeV}}$					&	$\Gamma_{1}$				&		$\Gamma_{2}$				&	$E_{\mathrm{break}}$		&		$-\log(\mathrm{Likelihood})$				&$\chi^{2}_{\mathrm{red}}$&	$\mathrm{TS}_{\mathrm{curve}}$ 	\\
							&	      	($10^{-6}$ ph cm$^{-2}$ s$^{-1}$)					&					&						&						(GeV)		&						&				\\
\hline
Pre-flare						&	1.40	$\pm$	0.15	&	1.76	 $\pm$	0.14		&	2.26 $\pm$ 0.18					&	0.9 $\pm$ 0.4				&	8758.38				&0.56&			7.00							\\
Flare						&	2.02	$\pm$	0.17	&	1.57  $\pm$	0.12			&	2.04 $\pm$ 0.14					&	1.3 $\pm$ 0.6				&	6285.55			&1.15&			6.98							\\
Post-flare 					&	1.06	$\pm$	0.12	&     1.72	 $\pm$	0.17			&	2.38 $\pm$ 0.22					&	1.0 $\pm$ 0.5				&	7734.79			&0.59&			6.48							\\
\hline
\end{tabular}
\label{Table3}
\end{table*}

\begin{table*}
\tablenum{4}
\begin{center}
\caption{Optimal Parameters of the Multiwavelength SED Modelings of PKS 1502+106 at Different Epochs of the $\gamma$-ray Flare in 2015.}
\scriptsize
\begin{tabular}{@{}ccccccccc@{}}
\hline\hline
Phase			&	$B$		&	$R_\mathrm{size}$	&				$p_{1}$			&	$\gamma_{\mathrm{br}}$		&	 $P_{r}$ 					&	$P_{B}$				         &$P_{e}$	& 	$P_{p}$ 					\\
				&	(G)		&	($10^{15} $ cm)	&		         					&	(10$^{3}$)				&	($10^{45}$ erg s$^{-1}$)		& 	($10^{44}$ erg s$^{-1}$)		&($10^{44}$ erg s$^{-1}$)	&	($10^{45}$ erg s$^{-1}$)		\\
(1)				&      (2)		&	(3)				&				(4)				&	(5)						&	 (6)						&	(7)						&(8)	&  	(9)						\\
\hline
Pre-flare				& 1.7 $\pm$ 0.6	&	6.3 $\pm$ 0.7	 &	2.23 $\pm$ 0.24	  & 1.2 $\pm$ 0.9 &	7.4	 & 	5.8	  & 	11.1	&	27.1	\\
Flare				& 1.0 $\pm$ 0.4	&	6.7 $\pm$ 0.8	 &	1.78 $\pm$ 0.28	  & 1.4 $\pm$ 0.8 &	13.3 & 	2.1	  & 	15.1	&	18.5	\\
Post-flare I			& 1.7 $\pm$ 0.5	&	3.6 $\pm$ 0.5	 &	1.07 $\pm$ 0.53	  & 0.4 $\pm$ 0.1 &	5.0	 & 	2.0	  & 	9.3 	&	7.9		\\
Post-flare II			& 1.7 $\pm$ 0.2	&	4.0 $\pm$ 0.4	 &	1.74 $\pm$ 0.19	  & 0.4 $\pm$ 0.1 &	5.5	 & 	2.4	  & 	10.6	&	16.3	\\
Post-flare III			& 1.6 $\pm$ 0.5	&	8.3 $\pm$ 1.5	 &	2.17 $\pm$ 0.29	  & 1.2 $\pm$ 1.0 &	5.9	 & 	9.8	  & 	6.5 	&	14.7	\\
\hline
\end{tabular}
\end{center}
\tablecomments{More physical meaning electronic kinetic power $P_{e}$ instead of electron number density $N_{0}$ is given in here.
The electronic kinetic power in AGN frame is calculated as $P_{e} = \pi R_{\mathrm{size}}^{2}\delta^{2}cm_{e}c^{2}\int_{\gamma_{\min}}^{\gamma_{\max}} \gamma N_{0}n(\gamma)\,d\gamma$.}
%
\label{Table4}
\end{table*}

\begin{figure}
\hspace{-0.9cm}
\includegraphics[width=3.8in]{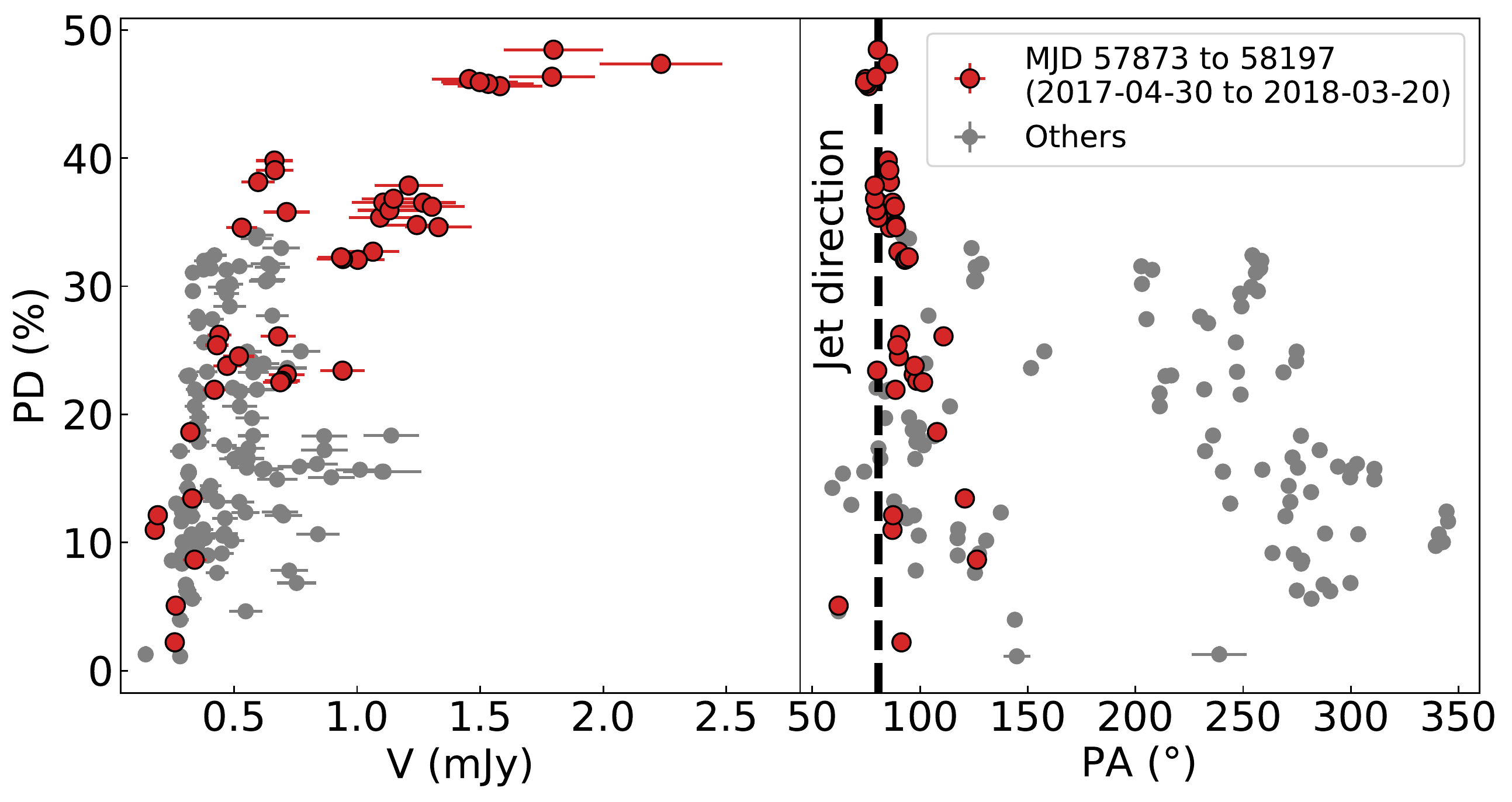}
\caption{The distributions of optical polarization degree versus optical V-band fluxes (left panel) and optical polarization angles (right panel). The data observed during the outburst in 2017 are marked in red. The black dashed line marks the parsec-scale jet position angle.}
\label{fig:correlation}
\end{figure}

\begin{figure}
\hspace{-0.5cm}
\includegraphics[width=3.2in]{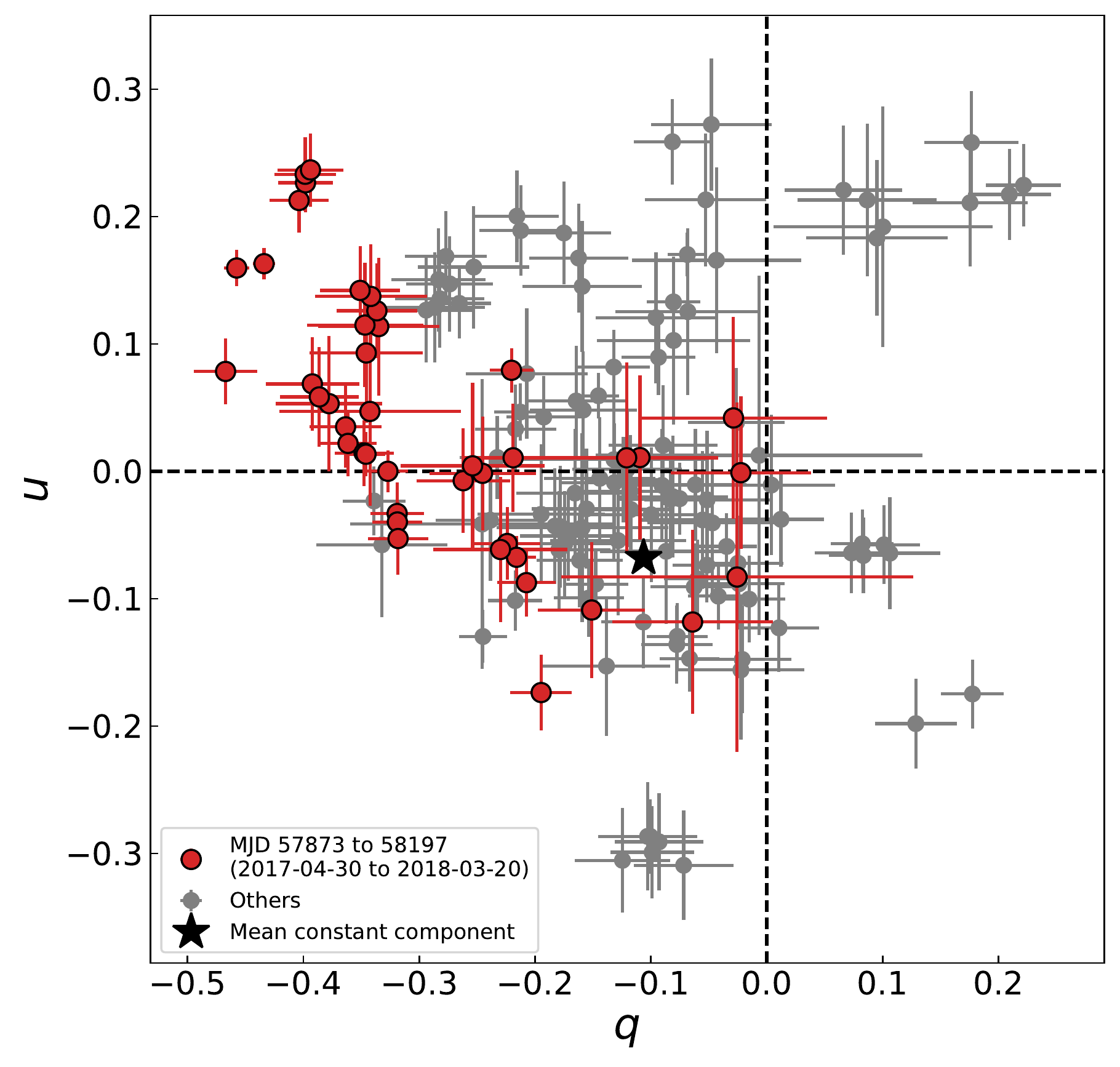}
\caption{Distribution plane of normalized Stokes parameters $q$ and $u$. The red data points represent the observed data during the outburst in 2017. The black star ($q_{c} = -0.106$ and $u_{c} = -0.068$) represents the average central point of $q$--$u$ calculated by iteratively discarding $>3 \sigma$ outliers.}
\label{fig:qu}
\end{figure}

\begin{figure}
\hspace{-0.5cm}
\includegraphics[width=3.2in]{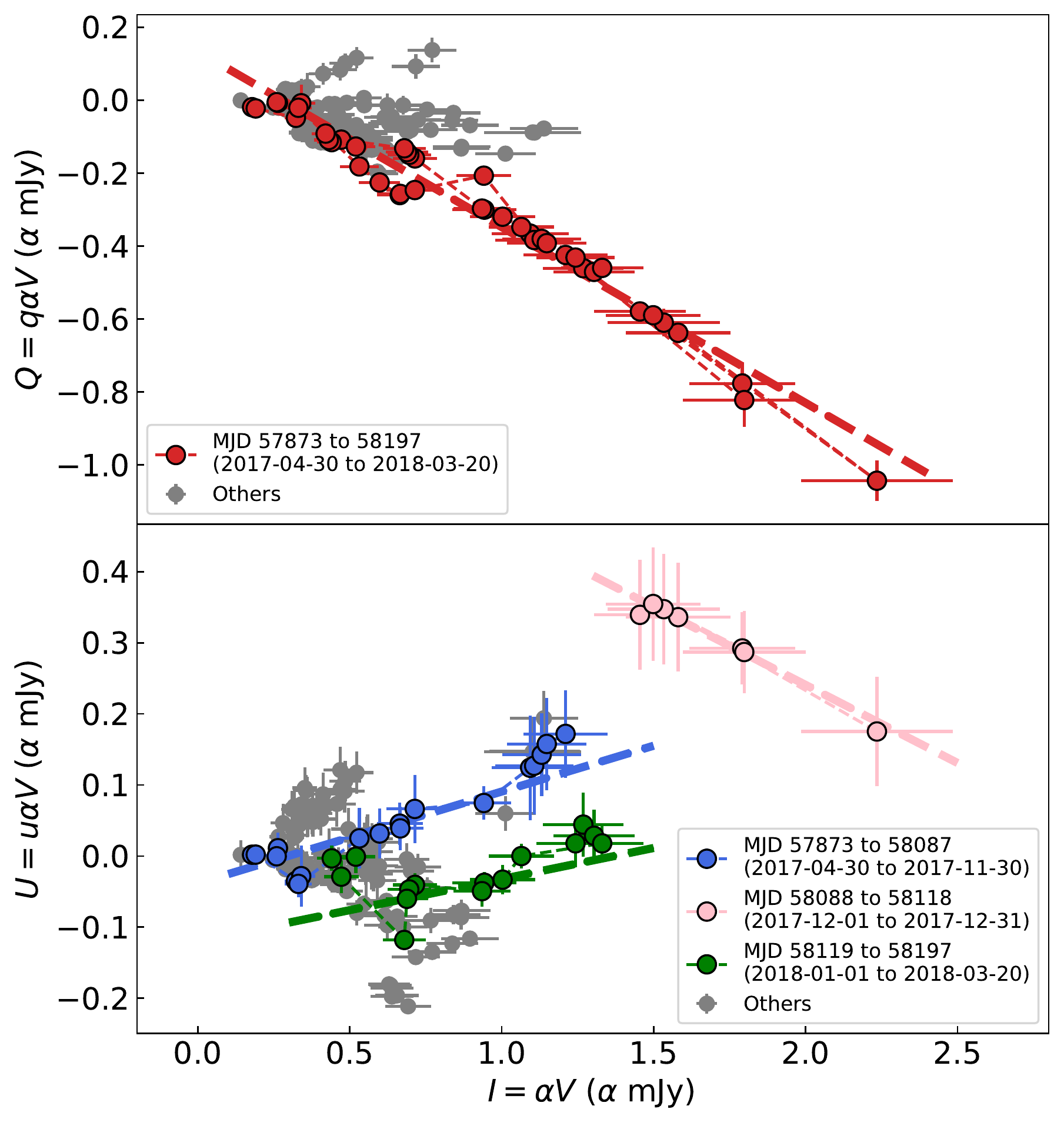}
\caption{Distribution planes of \mbox{$Q$--$I$} (top) and \mbox{$U$--$I$} (bottom). $(Q, U) = (q, u)I = (q, u)\alpha V$. The red data points in the top plane represent the observed data during the optical dominated outburst in 2017. The blue, pink, and green data points in the bottom plane correspond to the rising, peak and declining stages of the outburst, respectively. The colored dashed lines are the best linear fits for the corresponding color data points. The fitting results and the polarization parameters of the corresponding variable component are listed in Table~\ref{Table5}.}
\label{fig:UQI}
\end{figure}

\section{Optical dominated outburst in 2017}
As mentioned in Section~3.1, between May 2017 to March 2018, a significant optical dominant outburst occurred in PKS 1502+106. At the peak of the outburst, the long-term optical monitoring project at Steward Observatory monitored the highest optical polarization degree observed so far from PKS 1502+106, reaching $(47.4\pm0.1) \%$, which is also one of the highest polarization levels ever observed for blazars (ATel \#11047). 

The distributions of optical polarization degree versus optical V-band fluxes and optical polarization angles are shown in Figure~\ref{fig:correlation}. Here we use 1-day time interval to cross-match the data of photometry and polarization measurements, and finally 153 data points are obtained. The data observed during the outburst are marked in red. There is a very tight positive correlation between the optical polarization degree and the optical fluxes during the outburst, with a spearman correlation coefficient of 0.82 (the chance probability $P = 10^{-11}$). In addition, the optical polarization angles are within a narrow range during the outburst, with an average angle of $89.6^{\circ}$, which is close to the parsec-scale jet direction $\sim 81^{\circ}$ shown by the VLBI imaging \citep[][]{2016A&A...586A..60K}. 
In non-outburst state (grey data points), the optical polarization degree and the polarization angles exhibit random distributions (see also Figure~\ref{fig:qu}).

\citet[][]{2016A&A...590A..10K} used a simple random walk polarization variability model to analyze the long-term polarization observation data of FSRQ~3C279. They found that the polarization variation is possibly dominated by a stochastic process during the low-brightness state, while during the outburst, the polarization variation is governed by a deterministic process (e.g., shock compression, non-axisymmetric magnetic field configuration, etc). This is consistent with what we see in PKS~1502+106. During the outburst, the polarization behaviors of PKS 1502+106 are in agreement with the expectations of the \mbox{shock-in-jet} model, where the magnetic field is compressed and aligned at the front of the shock so that the polarization degree and fluxes exhibit a close positive correlation \citep[see, e.g.,][]{2008ApJ...672...40H}.
Interestingly, compared to outbursts with similar behaviors in other blazars discovered in previous studies (e.g., AO 0235+164 \citep{2008ApJ...672...40H}, 1ES 1959+650 \citep{2013ApJS..206...11S}), the outburst in PKS~1502+106 lasted for a long time ($\sim$ one year) and has extremely high polarization degree.

\subsection{Two-component decomposition}
The polarization observations of Steward Observatory provide normalized (relative) Stokes parameters $q$ and $u$. The \mbox{$q$--$u$} plane built from the data is shown in Figure~\ref{fig:qu}, where the red data points represent the observed data during the outburst and the black star ($q_{c} = -0.106$ and $u_{c} = -0.068$) represents the average central point of $q$--$u$ calculated by iteratively discarding $>3 \sigma$ outliers. The average central point are offset from the origin, and most of the red data points deviate from the average central point in the same direction. Therefore, we infer that the polarization of PKS~1502+106 could be composed of a constant or stable component associated with the jet configuration and a variable component that is related to the propagation of the shock \citep[][]{1984MNRAS.211..497H, 2008ApJ...672...40H}.

\begin{table*}
\tablenum{5}
\begin{center}
\caption{Polarization Properties of the Variable Component}
\scriptsize
\begin{tabular}{@{}ccccccc@{}}
\hline\hline
Stage							&	$R_{Q-I}$			&	$r_{Q-I}$				&	$R_{U-I}$						&	$r_{U-I}$			&			 		$P_{\mathrm{var}}$		&	$\Theta_{\mathrm{var}}$ 	\\
								&					&						&				 	&						&						($\%$)				&	($\deg$)			\\
(1)								&	(2)				&	(3)				&	(4)					&	(5)					&						(6)						&	(7)						\\
\hline
Rising						&								&					&		0.90			&	$0.13\pm0.02$						&	$49.7\pm1.4$							&	$82.4\pm1.3$			\\
\cline{1-1}
\cline{4-7}
Peak				&			-0.99				&	$-0.48\pm0.01$    	&		-0.89		&	$-0.22\pm0.02$						&	$52.8\pm1.7$							&		$102.3\pm1.2$		\\
\cline{1-1}
\cline{4-7}
Declining						&								&     					&		0.59			&	$0.08\pm0.05$						&	$48.6\pm1.8$							&	$85.3\pm3.0$			\\
\hline
\end{tabular}
\end{center}
\tablecomments{Columns from left to right: (1) different stages of the outburst. (2) correlation coefficient of $Q$--$I$ relation. (3) slope of $Q$--$I$ linear fitting. (4) correlation coefficient of $U$--$I$ relation. (5) slope of $U$--$I$ linear fitting. 
(6) polarization degree of corresponding variable component ($P_{\mathrm{var}} = \sqrt{q_{var}^{2}+u_{var}^{2}}$). (7) polarization angle of corresponding variable component ($\Theta_{var} = \frac{1}{2}\arctan(u_{var}/q_{var})$).}
\label{Table5}
\end{table*}

Based on the average central point, the polarization degree and polarization angle of the stable component are calculated as $P_{c} = \sqrt{q_{c}^{2}+u_{c}^{2}} = 12.6\% $  and $\Theta_{c} = \frac{1}{2}\arctan(u_{c}/q_{c}) = 106.3^{\circ}$, respectively. For the variable component, following the method proposed by \citet{2008ApJ...672...40H}, we use absolute Stokes parameter planes, i.e., \mbox{$Q$--$I$} and \mbox{$U$--$I$} planes, to determine its variability behaviors, where the absolute Stokes parameters are defined as $(Q, U) = (q, u)I$. If the variability of the variable composition is only caused by its flux variations, independent of its polarization degree and polarization angle, the relative Stokes parameters ($q_v$ and $u_v$) of the variable composition will remain unchanged. 
In this case, observed data points will lie on straight lines in the absolute Stokes parameter planes ($Q$--$I$ and $U$--$I$ planes), and the slopes ($r_{Q-I}$ and $r_{U-I}$) of these lines reflect the relative Stokes parameters of the variable component (i.e., $q_{var} = r_{Q-I}$ and $u_{var} = r_{U-I}$). 
The \mbox{$Q$--$I$} and \mbox{$U$--$I$} planes built from our observed data are shown in Figure~\ref{fig:UQI}, where $\alpha$ is a proportional term between $I$ and the V-band fluxes. In the $Q$--$I$ plane, the data points during the outburst (red points) show a very tight linear dependence, indicating that there is no change in the relative Stokes parameter $q_v$ during the outburst. In the $U$--$I$ plane, the data points during the outburst display three-stage linear relations. The blue, pink, and green data points correspond to the rising, peak and declining stages of the outburst, respectively. 
We perform linear fittings separately for the $U$--$I$ relations at the three different stages while performing only one linear fitting for $Q$--$I$ relation. The fitting results and the polarization parameters of the corresponding variable component are listed in Table~\ref{Table5}. The variable component shows extremely high polarization degree, up to ($52.8\pm1.7)\%$ in the peak stage. The polarization angle of the variable component is close to the parsec-scale jet direction at the rising and declining stages, and it has a change of $\sim20^{\circ}$ at the peak stages.

\begin{figure}
\hspace{-0.5cm}
\includegraphics[width=3.2in]{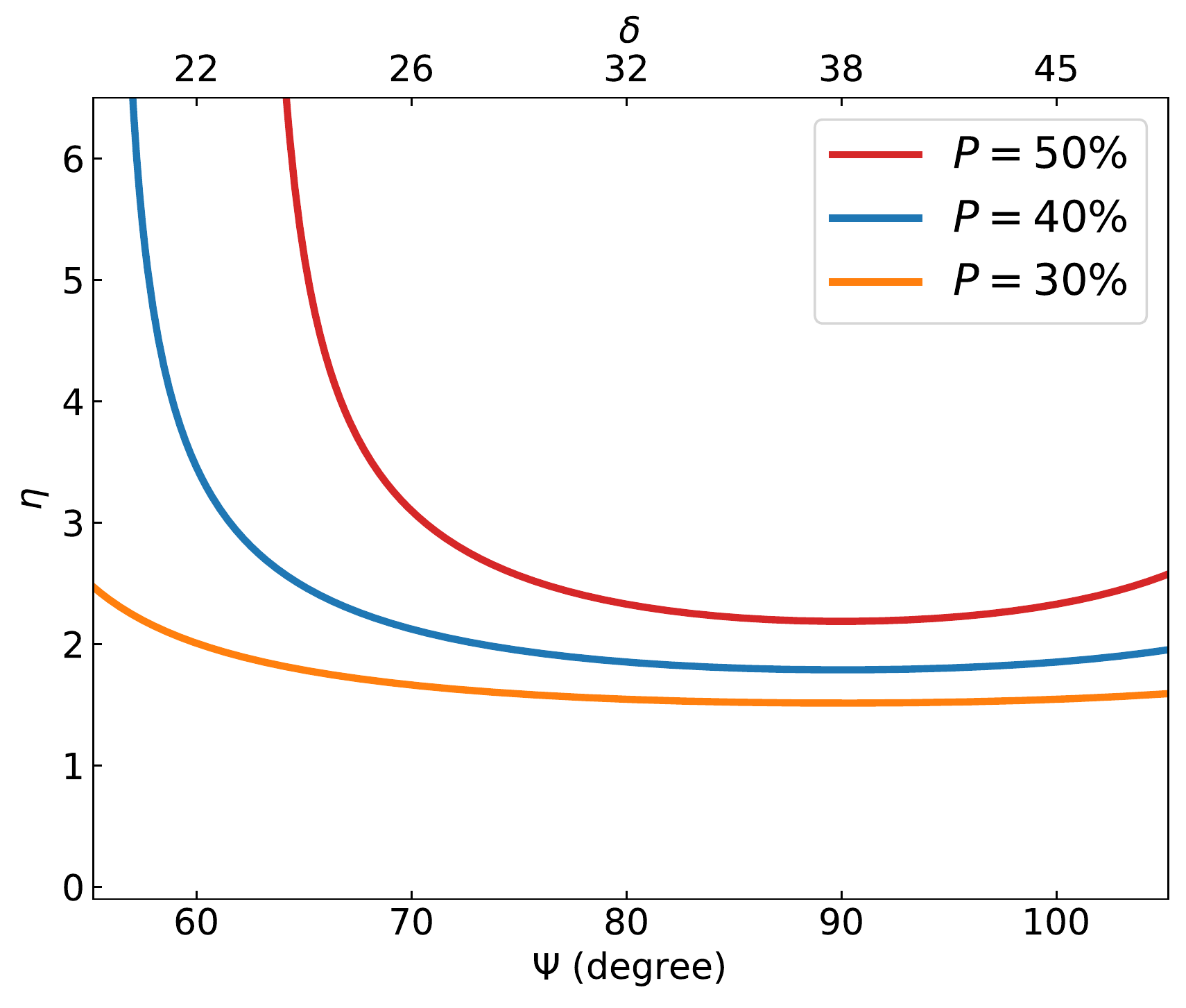}
\caption{The functions of the viewing angle of the shock and the compression ratio under the different polarization degrees. The values of the Doppler factor corresponding to the viewing angles of the shock are also marked in the figure.}
\label{fig:Psi_eta}
\end{figure}

\begin{figure}
\hspace{-0.5cm}
\includegraphics[width=3.2in]{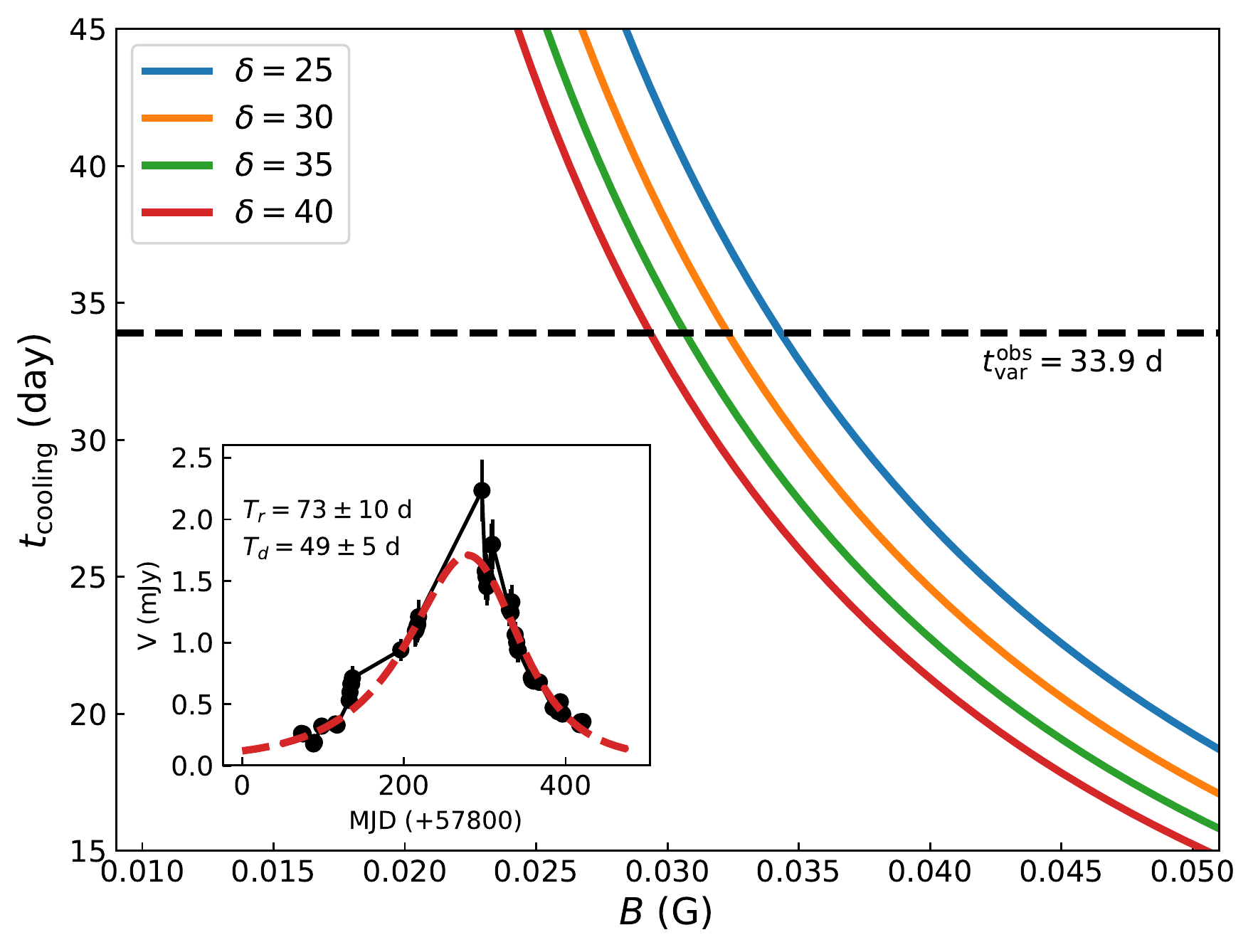}
\caption{The functions of the magnetic field intensity and the cooling time under the different Doppler factors. The time-profile fitting of the outburst in the V band is shown in the inset. The black line represents the shortest doubling/halving time-scale. }
\label{fig:tB}
\end{figure}

\subsection{The properties of the shock}
Based on the variability behaviors seen above, below we briefly discuss the properties of the shock. As observed, this dramatic outburst seen in PKS 1502+106 occurred mainly in the optical band, suggesting that the emission in the shock zone is dominated by synchrotron radiation. In addition, the polarization angle of the variable component is close to the parsec-scale jet direction, which implies that the shock could be a transverse shock; the transverse shock orders the turbulent magnetic field along the front, which is perpendicular to the jet direction \citep{2008ApJ...672...40H}. In this scenario, the polarization degree depends on the viewing angle of the shock $\Psi$, the spectral index $\alpha_{\mathrm{o}}$ in optical band, and the ratio of densities of the shocked region to the unshocked region (compression ratio) $\eta = n_{\mathrm{shock}}/n_{\mathrm{unshock}}$ \citep[][]{1991bja..book....1H}:
\begin{equation}
\label{eq6}
p \approx \frac{\alpha+1}{\alpha+5/3} \frac{(1-\eta^{-2})\sin^{2}\Psi}{2-(1-\eta^{-2})\sin^{2}\Psi}.
\end{equation}
The viewing angle of the shock in the observer's frame is subjected to relativistic aberration and determined by the bulk Lorentz factor $\Gamma$ and the viewing angle of the jet $\theta$:
\begin{equation}
\label{eq7}
\Psi = \tan^{-1}\{\sin \theta/\Gamma(\cos \theta-\sqrt{1-\Gamma^{-2}})\}.
\end{equation}

According to Eq.~\ref{eq5}, the functions of the viewing angle of the shock and the compression ratio under the different polarization degrees are displayed in Figure~\ref{fig:Psi_eta}. Meanwhile, according to Eq.~\ref{eq6} and the relation of $\delta = [\Gamma(1-(1-\Gamma^{-2})^{1/2}\cos\theta)]^{-1}$, the values of the Doppler factor corresponding to the viewing angles of the shock are marked in the figure. Here, we use the value of $\alpha_{\mathrm{o}} = 1.8$ calculated from the average optical data during the outburst and the value of $\theta \approx 3^{\circ}$ constrained by the mm-VLBI observations \citep[][]{2016A&A...586A..60K}. It can be seen that in a wide range of the Doppler factor, the polarization degree of the variable component wants to reach observed $\sim50\%$, the compression ratio of the shock must be no less than $2.2$.

The time-scale of outburst relates to the thickness of the shock front, which is determined by the lifetime of relativistic electrons accelerated at the front. In the current situation dominated by synchrotron radiation, the lifetime of relativistic electrons in the observer frame is
\begin{equation}
\label{eq8}
t_{\mathrm{cooling}} \approx 4.75\times10^{2} \frac{(1+z)} {\delta\nu_{\mathrm{GHz}}B^{3}}~~\mathrm{days},
\end{equation}
where $\nu_{\mathrm{GHz}}$ is observed synchrotron photon frequency in GHz \citep[][]{2008ApJ...672...40H}. The functions of the magnetic field intensity and the cooling time under the different Doppler factors are displayed in Figure~\ref{fig:tB}. As in Section 4.1, we use a double exponential function to fit the time profile of the outburst in the V band. The fitting result is shown in the inset of Figure~\ref{fig:tB}. The rising and decaying time-scales of this outburst are $T_{r} = (73\pm 10)$~d and $T_{d} = (49\pm 5)$~d, and the corresponding fastest variability time (i.e., shortest doubling/halving time-scale) is $t_{\mathrm{var}}^{\mathrm{obs}} = (31.9\pm 2.7)$~d (marked as black line in Figure~\ref{fig:tB}). 
In the case where the cooling time is equivalent to the fastest variability time, the magnetic field intensity is required to be $\sim0.032$~G and weakly depends on the Doppler factor. 
\citet{2016A&A...590A..48K} performed a cross-correlation analysis for the radio light curves observed during the $\gamma$-ray outburst in 2008. Based on the observed time-delays between different bands, the structure of PKS 1502+106 in terms of synchrotron opacity was deduced, while the magnetic field intensity along the jet axis also was estimated (under the \mbox{shock-in-jet} scenario). They estimated the magnetic field intensity in the radio nucleus to be between 14 and 176~mG. Our estimated magnetic field intensity in the shock emission region is in harmony with their result. However, this magnetic field intensity is much smaller than that obtained in the $\gamma$-ray flare in 2015, which implies that the properties of the emission regions of the activities triggered by shock and magnetic reconnection are quite different.

\section{Summary}
After the $\gamma$-ray activities in 2008--2009, PKS 1502+106 entered a quiescent stage of up to six years. Until mid-2015, its prominent multi-band activities are re-detected. Using multi-band data from radio to $\gamma$-ray bands as well as optical polarization observations, we systematically explore the multiwavelength activities of high-redshift FSRQ PKS 1502+106 during 2014--2018. Two dramatic outbursts, a $\gamma$-ray dominated outburst in 2015 and an optical dominated outburst in 2017, are investigated in detail to explore the triggering mechanism of them and the physical properties of the emission regions. The main results are summarized as follows.

1. An hour-scale GeV $\gamma$-ray flare is discovered during the $\gamma$-ray dominated outburst in 2015. This fast flare shows obviously asymmetric time profile, and the $\gamma-$ray spectral index is $\Gamma_{\gamma} = 1.82\pm0.04$ at the peak of the flare, which is rarely seen in FSRQs. Based on the variability time-scale of the flare, the physical parameters of flaring region (e.g, minimum Doppler factor, emission region size, etc.) are constrained. See~Section~4.1. 

2. The $\gamma$-ray spectra at the different phases of the flare can be best fitted by the PLEC function. At the Flare phase, $\gamma$-ray spectrum emerges a curvature/break characteristic ($\sim3.2\sigma$). The curvature/break characteristic is in line with the expectation of typical "cooling break", suggesting that it is most likely caused by the radiative losses of high-energy electrons. See~section~4.2.

3. Based on a one-zone homogeneous leptonic model, multi-band SEDs at different epochs of the $\gamma$-ray flare are modeled. The results show that the multi-band radiation of PKS 1502+106 needs the involvement of EC process, and the soft photons in the EC process should mainly come from the BLR, which further confirms the result in \citet{2010ApJ...710..810A}. In addition, SED modelings reveal the changes of the electron spectral index and magnetic field intensity in the emission region at different epochs of the flare. 
In the Flare phase, the magnetic field intensity seems to decrease slightly, and then a harder electron spectrum with the electronic spectral index of $p_{1} = 1.07\pm0.53$ appears. This result may imply that the fast $\gamma$-ray flare is generated by magnetic reconnection. See~section~4.3.

The above conclusions are obtained in the framework of the leptonic scenario. In view of typical limitations in the \mbox{lepto-hadronic} models (see Section~4.3), the \mbox{lepto-hadronic} models are not further explored in our work. 
Nevertheless, it should be noted that the \mbox{lepto-hadronic} scenario may still a potential alternative.
For example, some studies on the giant $\gamma$-ray flare of 3C 279 in June 2015 (similar to the time scale of the flare in PKS 1502+106) show that the lepto-hadronic scenario still has the potential to explain such fast $\gamma$-ray flare under certain conditions \citep[see, e.g.,][]{2017MNRAS.467L..16P, 2019arXiv190604996H}. Whether the emission origin of blazars is leptonic scenario or lepto-hadronic scenario is still a controversial issue, which needs to be further explored in the future.

4. An optical dominated outburst occurred in 2017. During the outburst, the optical polarization degree and optical fluxes exhibit a very significant correlation. By analyzing Stokes parameters of polarization observations, our results show that the observed polarization could be composed of a stable component associated with the jet configuration and a variable component that is related to the propagation of shock. The polarization degree of the variable component is as high as $(52.8\pm1.7)\%$ at the peak of the outburst. The outburst could be triggered by a transverse shock with a compression ratio of $\eta>2.2$, and the magnetic field intensity of the shock emission region is about $0.032$~G. See~Section~5.

In PKS 1502+106, we see that both shock and magnetic reconnection may be the triggers of multi-band activities, and the multi-band activities triggered by the two may be significantly different. Perhaps the short-time high-energy flares are more likely to be triggered by magnetic reconnection, while the \mbox{long-term} outbursts in a low-energy band dominated by synchrotron radiation are more likely to be correlated with shocks. 
At present, there are few reports of flares triggered by magnetic reconnection. This may be attributed to the need for multi-band synergetic observation data to identify such flares. 
In addition, such flares perhaps have short time-scale, which makes it rare for such events to have multi-band simultaneous observations, thereby reducing the recognition rate of such flares. 
The coming era of multi-messenger time-domain astronomy will provide more high-quality multi-band synergetic observation data. By then, the studies of the multi-band activities for blazars will reveal a more complete picture of the energy dissipation mechanism in the jet \citep[see, e.g.,][]{2019arXiv190304504R, 2019arXiv190304461B}.
\\
\acknowledgments
We sincerely thank the anonymous referee for helpful suggestions.
We acknowledge financial support from the National Key R\&D Program of China grant 2017YFA0402703 (N.D., Q.S.Gu) and National Natural Science Foundation of China grant 11733002 (N.D., Q.S.Gu).
Dingrong Xiong acknowledges financial support from the National Natural Science Foundation of China grant 11703078.

This work has made use of Fermi data, obtained from the Fermi Science Support Center, provided by NASA's Goddard Space Flight Center (GSFC). 
The data, software, and web tools obtained from NASA's High Energy Astrophysics Science Archive Research Center (HEASARC), a service of GSFC, were used. 
This work has made use of the Swift and XMM-Newton data. Also, data from the Steward Observatory spectropolarimetric monitoring project were used. This program is supported by Fermi Guest Investigator grants NNX08AW56G, NNX09AU10G, NNX12AO93G, and NNX15AU81G. 
We also has made use of data from the OVRO 40 m monitoring program, which is supported in part by NASA grants NNX08AW31G, NNX11A043G, and NNX14AQ89G and NSF grants AST-0808050 and AST-1109911.

\software{Fermi Science Tools \citep{2014AAS...22325528A}, HEAsoft \citep{2014ascl.soft08004N}, SAS \citep{2004ASPC..314..759G}, XSPEC \citep{1996ASPC..101...17A}, 
LMFIT \citep{2016ascl.soft06014N}, Astropy \citep{2013A&A...558A..33A}, SciPy \citep{Jones et al.(2001)}, NumPy \citep{Walt et al.(2011)}, Matplotlib \citep{2005ASPC..347...91B}}

\end{document}